\documentclass[12pt]{article}

\usepackage{amsmath}
\usepackage{amssymb}
\usepackage{bbm}
\usepackage{hyperref}
\usepackage{feynmp-auto}
\usepackage{physics}
\usepackage{graphicx}
\usepackage{feynmp-auto}
\usepackage{subcaption}

\newcommand{\be}{\begin{equation}}
\newcommand{\ee}{\end{equation}}
\newcommand{\ba}{\begin{eqnarray}}
\newcommand{\ea}{\end{eqnarray}}
\newcommand{\bs}{\begin{subequations}}
\newcommand{\es}{\end{subequations}}
\newcommand{\no}{\nonumber\\}

\newcommand{\N}{\mathcal{N}}
\newcommand{\V}{\mathcal{V}}
\newcommand{\Q}{\mathcal{Q}}
\newcommand{\Hh}{\mathcal{H}}
\newcommand{\Uu}{\mathcal{U}}
\newcommand{\Dd}{\mathcal{D}}
\newcommand{\Ll}{\mathcal{L}}

\newcommand{\marrow}[6]{
    \fmfcmd{style_def marrow#1
    expr p = drawarrow subpath #6 of p shifted 10 #2 withpen pencircle scaled 0.5; label.#3(btex #4 etex, point 0.5 of p shifted 12 #2);
    enddef;}
    \fmf{marrow#1,tension=0}{#5}}

\begin{document}

\begin{fmffile}{diagram}

\title{\LARGE Oblique corrections from leptoquarks}

\author{\addtocounter{footnote}{2}
  Francisco Albergaria\thanks{E-mail:
  \href{mailto:francisco.albergaria@tecnico.ulisboa.pt}{\tt francisco.albergaria@tecnico.ulisboa.pt}}
  \ and\addtocounter{footnote}{2}
  Lu\'\i s Lavoura\thanks{E-mail: \href{mailto:balio@cftp.tecnico.ulisboa.pt}{\tt balio@cftp.tecnico.ulisboa.pt}}
  \\*[3mm]
  \small Universidade de Lisboa, Instituto Superior T\'ecnico, CFTP, \\
  \small Av.~Rovisco Pais~1, 1049-001 Lisboa, Portugal
  \\*[2mm]}

\date{\today}

\maketitle

\begin{abstract}
  We present general formulas for the oblique-correction parameters $S$,
  $T$,
  $U$,
  $V$,
  $W$,
  and $X$ in a model of New Physics
  having arbitrary numbers of scalar leptoquarks
  of the five permissible types.
  We allow for a general mixing among the scalars
  of the various electric charges,
  \textit{viz.}\ $-4/3$,
  $-1/3$,
  $2/3$,
  and $5/3$.
  We then extend the formulas to the case of a New Physics model
  with additional scalars in any representations
  of the gauge group $SU(2) \times U(1)$,
  mixing arbitrarily among themselves.
\end{abstract}

\vspace*{4mm}

\section{Introduction and notation}

\paragraph{Leptoquarks:} In this paper we consider a model of New Physics (NP),
\textit{i.e.}\ an extension of the Standard Model (SM),
that has arbitrary numbers of scalars
placed in triplets of colour-$SU(3)$ which are
\begin{description}
\item $SU(2)$ singlets
  with weak hypercharge\footnote{We use the normalization $Y = Q - T_3$,
  where $Y$ is the weak hypercharge,
  $Q$ is the electric charge,
  and $T_3$ is the third component of weak isospin.} $-1/3$
  \be
  \sigma_{0,-2};
  \label{a}
  \ee
\item $SU(2)$ singlets with weak hypercharge $-4/3$
  \be
  \sigma_{0,-8};
  \label{b}
  \ee
\item $SU(2)$ doublets with weak hypercharge $7/6$
  \be
  \left( \begin{array}{c} \delta_{1,7} \\
    \delta_{-1,7} \end{array} \right);
  \label{c}
  \ee
\item $SU(2)$ doublets with weak hypercharge $1/6$
  \be
  \left( \begin{array}{c} \delta_{1,1} \\
    \delta_{-1,1} \end{array} \right);
  \label{d}
  \ee
\item $SU(2)$ triplets with weak hypercharge $-1/3$
  \be
  \left( \begin{array}{c} \tau_{2,-2} \\ \tau_{0,-2} \\
    \tau_{-2,-2} \end{array} \right).
  \label{g}
  \ee
\end{description}
In Eqs.~\eqref{a}--\eqref{g},
\begin{description}
\item the letter $\sigma$ denotes singlets of gauge $SU(2)$,
  the letter $\delta$ stands for doublets,
  and the letter $\tau$ means triplets;
\item the first number in the subscript is
  two times the third component of weak isospin $T_3$;
\item the second number in the subscript is
  six times the weak hypercharge $Y$.
\end{description}
In the notation of the recent Ref.~\cite{crivellin},
which refers to the original Ref.~\cite{buch},
\begin{description}
\item the scalars in Eq.~\eqref{a} are leptoquarks of type $\Phi_1$,
  \textit{viz.}\ scalars placed in the representation
  $\left( \mathbf{3}, \mathbf{1}, -1/3 \right)$
  of the gauge group $SU(3) \times SU(2) \times U(1)$;
\item the scalars in Eq.~\eqref{b} are leptoquarks of type $\Phi_{\tilde 1}$,
  \textit{viz.}\ scalars placed in the representation
  $\left( \mathbf{3}, \mathbf{1}, -4/3 \right)$
  of $SU(3) \times SU(2) \times U(1)$;
\item the scalars in Eq.~\eqref{c} are leptoquarks of type $\Phi_2$,
  \textit{viz.}\ scalars placed in the representation
  $\left( \mathbf{3}, \mathbf{2}, 7/6 \right)$
  of the gauge group;
\item the scalars in Eq.~\eqref{d} are leptoquarks of type $\Phi_{\tilde 2}$,
  \textit{viz.}\ scalars placed in the representation
  $\left( \mathbf{3}, \mathbf{2}, 1/6 \right)$;
\item the scalars in Eq.~\eqref{g} are leptoquarks of type $\Phi_3$,
  \textit{viz.}\ scalars placed in the representation
  $\left( \mathbf{3}, \mathbf{3}, -1/3 \right)$
  of $SU(3) \times SU(2) \times U(1)$.
\end{description}
Leptoquarks\footnote{In this paper we focus exclusively
on \emph{scalar} leptoquarks.
Various authors use the term `leptoquarks' to mean some vector,
\textit{i.e.}\ spin-one,
fields.}
are scalars
that are in multiplets of $SU(3) \times SU(2) \times U(1)$
such that they have (renormalizable) Yukawa couplings
to one lepton multiplet and one quark multiplet of the SM.
Leptoquarks have recently been much used in models
that seek to explain one or more unexpected experimental results
like those on the muon magnetic moment,
the $W^\pm$ mass,
the decays $b \to c \tau \nu$,
and the decays $b \to s \ell^+ \ell^-$;
see for instance Refs.~\cite{ks1,ks2,ks3,ks4,ks5,ks6,ks7,ks8}.

\paragraph{Oblique parameters:}
The oblique parameters are defined as~\cite{maksymyk}\footnote{We use
the sign conventions in Ref.~\cite{book}.
Those conventions differ from the ones used in many other papers,
\textit{viz.}\ in Ref.~\cite{maksymyk}.
For a resource paper on sign conventions,
see Ref.~\cite{romao1};
using the notation of that paper,
our convention has $\eta_e = \eta_Z = 1$ and $\eta = -1$.}$^,$\footnote{The
definitions~\eqref{jbifgopdde} build on,
and generalize,
previous work in Refs.~\cite{peskin,altarelli,indianos}.
They are appropriate for the case where the functions
$A_{V V^\prime} \left( q^2 \right)$ are not linear in the range $0 < q^2 < m_Z^2$.
This means that,
using those definitions,
NP does not need to be much above the Fermi scale.}
\bs
\label{jbifgopdde}
\ba
S &=& \frac{16 \pi c_W^2}{g^2} \left[
  \frac{A_{ZZ} \left( m_Z^2 \right) - A_{ZZ} \left( 0 \right)}{m_Z^2}
  - \left. \frac{\partial A_{\gamma\gamma} \left( q^2 \right)}{\partial q^2}
  \right|_{q^2 = 0}
  \right. \no & & \left.
  + \frac{c_W^2 - s_W^2}{c_W s_W}\,
  \left. \frac{\partial A_{\gamma Z} \left( q^2 \right)}{\partial q^2}
  \right|_{q^2 = 0} \right],
\\
T &=& \frac{4 \pi}{g^2 s_W^2} \left[
  \frac{A_{WW} \left( 0 \right)}{m_W^2}
  - \frac{A_{ZZ} \left( 0 \right)}{m_Z^2} \right],
\\
U &=& \frac{16 \pi}{g^2} \left[
  \frac{A_{WW} \left( m_W^2 \right) - A_{WW} \left( 0 \right)}{m_W^2}
  - c_W^2\, \frac{A_{ZZ} \left( m_Z^2 \right) - A_{ZZ} \left( 0 \right)}{m_Z^2}
  \right. \no & & \left.
  - s_W^2 \left. \frac{\partial A_{\gamma \gamma} \left( q^2 \right)}{\partial q^2}
  \right|_{q^2 = 0}
  + 2 c_W s_W \left. \frac{\partial A_{\gamma Z} \left( q^2 \right)}{\partial q^2}
  \right|_{q^2 = 0}
  \right],
\label{Udef} \\
V &=& \frac{4 \pi}{g^2 s_W^2} \left[
  \left. \frac{\partial A_{ZZ} \left( q^2 \right)}{\partial q^2}
  \right|_{q^2 = m_Z^2}
  - \frac{A_{ZZ} \left( m_Z^2 \right) - A_{ZZ} \left( 0 \right)}{m_Z^2}
  \right],
\\
W &=& \frac{4 \pi}{g^2 s_W^2} \left[
  \left. \frac{\partial A_{WW} \left( q^2 \right)}{\partial q^2}
  \right|_{q^2 = m_W^2}
  - \frac{A_{WW} \left( m_W^2 \right) - A_{WW} \left( 0 \right)}{m_W^2}
  \right],
\\
X &=& \frac{4 \pi c_W}{g^2 s_W} \left[
  \left. \frac{\partial A_{\gamma Z} \left( q^2 \right)}{\partial q^2}
  \right|_{q^2 = 0} -
  \frac{A_{\gamma Z} \left( m_Z^2 \right) - A_{\gamma Z} \left( 0 \right)}{m_Z^2}
  \right],
\label{Xdef}
\ea
\es
where $g$ is the $SU(2)$ gauge coupling constant
and $c_W$ and $s_W$ are the cosine and the sine,
respectively,
of the Weinberg angle.
The functions $A_{V V^\prime} \left( q^2 \right)$
are the coefficients of the metric tensor $g^{\mu \nu}$
in the vacuum-polarization tensor
\be
\Pi^{\mu \nu}_{V V^\prime} \left( q^2 \right)
= g^{\mu \nu}\, A_{V V^\prime} \left( q^2 \right)
+ q^\mu q^\nu\, B_{V V^\prime} \left( q^2 \right)
\ee
between gauge bosons $V_\mu$ and $V^\prime_\nu$ carrying four-momentum $q$.
In $A_{V V^\prime} \left( q^2 \right)$
\begin{description}
\item one only takes into account the dispersive part---one discards
  the absorptive part;
\item one subtracts the SM contribution
  from the full result.\footnote{In this paper
  we do not need to perform this subtraction,
  since we are dealing with NP scalars that
  have non-integer electric charges and,
  therefore,
  do not mix with the SM scalars.}
\end{description}
Note that
\be
S + U = \frac{16 \pi}{g^2} \left[
  \frac{A_{WW} \left( m_W^2 \right) - A_{WW} \left( 0 \right)}{m_W^2}
  - \left. \frac{\partial A_{\gamma \gamma} \left( q^2 \right)}{\partial q^2}
  \right|_{q^2 = 0}
  + \frac{c_W}{s_W}
  \left. \frac{\partial A_{\gamma Z} \left( q^2 \right)}{\partial q^2}
  \right|_{q^2 = 0}
  \right]
\label{SUdef}
\ee
has a somewhat simpler expression than $U$.

It is convenient to separate the parameters $S$ and $U$ in two parts:
\be
\label{dec1}
S = S^\prime + S^{\prime \prime}, \quad U = U^\prime + U^{\prime \prime}.
\ee
The parameters $S^\prime$ and $U^\prime$
are identical to the original $S$ and $U$,
respectively,
as they were defined in Ref.~\cite{peskin}:
\bs
\label{dec2}
\ba
S^\prime &=& \frac{16 \pi c_W^2}{g^2} \left[
  \left. \frac{\partial A_{ZZ}
    \left( q^2 \right)}{\partial q^2} \right|_{q^2 = 0}
  - \left. \frac{\partial A_{\gamma \gamma}
    \left( q^2 \right)}{\partial q^2} \right|_{q^2 = 0}
  + \frac{c_W^2 - s_W^2}{c_W s_W} \left. \frac{\partial A_{\gamma Z}
    \left( q^2 \right)}{\partial q^2} \right|_{q^2 = 0}
  \right],
\hspace*{7mm} \\
U^\prime &=& - S^\prime + \frac{16 \pi}{g^2} \left[
  \left. \frac{\partial A_{WW}
    \left( q^2 \right)}{\partial q^2} \right|_{q^2 = 0}
  - \left. \frac{\partial A_{\gamma \gamma}
    \left( q^2 \right)}{\partial q^2} \right|_{q^2 = 0}
  + \frac{c_W}{s_W} \left. \frac{\partial A_{\gamma Z}
    \left( q^2 \right)}{\partial q^2} \right|_{q^2 = 0}
  \right].
\ea
\es
Clearly then,
\bs
\label{dec3}
\ba
S^{\prime \prime} &=& \frac{16 \pi c_W^2}{g^2} \left[
  \frac{A_{ZZ} \left( m_Z^2 \right) - A_{ZZ} \left( 0 \right)}{m_Z^2}
  - \left. \frac{\partial A_{ZZ}
    \left( q^2 \right)}{\partial q^2} \right|_{q^2 = 0}
  \right],
\\
U^{\prime \prime} &=& - S^{\prime \prime} + \frac{16 \pi}{g^2} \left[
  \frac{A_{WW} \left( m_W^2 \right) - A_{WW} \left( 0 \right)}{m_W^2}
  - \left. \frac{\partial A_{WW}
    \left( q^2 \right)}{\partial q^2} \right|_{q^2 = 0}
  \right].
\ea
\es

\paragraph{Purpose of this paper:}
In this paper we compute the oblique parameters for a NP model
with arbitrary numbers of leptoquarks,
mixing arbitrarily among themselves.
Our work generalizes earlier partial results
in Refs.~\cite{earl1,earl2,earl3,earl4,earl5,earl6}.
At the end of this paper we present a generalization
wherein the oblique parameters are computed for any NP model
solely with extra scalars,
whatever the representations of $SU(2) \times U(1)$ those new scalars
are in---just assuming that no new scalar
either develops a vacuum expectation value or mixes with the scalars of the SM.

\paragraph{Plan of the paper:}
In the next section we write the Lagrangian
for the gauge interactions of the scalars,
carefully defining the mixing matrices that appear in that Lagrangian.
Section~3 performs the computation of the relevant diagrams
and explicitly gives various functions that appear in the formulas
for the oblique parameters.
Those formulas are then given in Section~4.
Section~5 presents a few illustrative examples of application
of our formulas to simple models with leptoquarks.
Section~6 generalizes our work to scalars in any representations
of $SU(2) \times U(1)$.
Section~7 summarizes our main results.
In Appendices~A and~B we deal,
respectively,
with the counting of mixing parameters
and with the parameterization of the mixing in models with leptoquarks.
In Appendix~C we explicitly demonstrate that $S^\prime$ and $U^\prime$
are finite as they should be.

\section{Interactions}

\paragraph{Numbers of scalars:} There are in our NP model
$n_{\sigma,-2}$ multiplets of type~\eqref{a},
$n_{\sigma,-8}$ multiplets of type~\eqref{b},
$n_{\delta,7}$ multiplets of type~\eqref{c},
$n_{\delta,1}$ multiplets of type~\eqref{d},
and $n_{\tau,-2}$ multiplets of type~\eqref{g}.
All these five numbers $n_{\sigma,-2}$,
$n_{\sigma,-8}$,
$n_{\delta,7}$,
$n_{\delta,1}$,
and $n_{\tau,-2}$ must be multiples of three,
since all the scalars are in triplets of $SU(3)$;
otherwise,
those numbers are free.

Our model has fractionary-charge scalars that we call
\begin{description}
\item $h$-type scalars,
  \textit{viz.}\ the ones that have electric charge $Q_h = 5/3$;
\item $u$-type scalars,
  \textit{viz.}\ the ones that have electric charge $Q_u = 2/3$;
\item $d$-type scalars,
  \textit{viz.}\ the ones that have electric charge $Q_d = - 1/3$;
\item $l$-type scalars,
  \textit{viz.}\ the ones that have electric charge $Q_l = - 4/3$.
\end{description}
The total numbers of $h$-type scalars,
$u$-type scalars,
$d$-type scalars,
and $l$-type scalars are
\bs
\label{tri}
\ba
n_h &=& n_{\delta,7},
\\
n_u &=& n_{\delta,7} + n_{\delta,1} + n_{\tau,-2},
\\
n_d &=& n_{\sigma,-2} + n_{\delta,1} + n_{\tau,-2},
\\
n_l &=& n_{\sigma,-8} + n_{\tau,-2},
\ea
\es
respectively.
Notice that
\be
n_d \ge n_u - n_h \ge 0.
\ee

\paragraph{Scalar mixing:}
One bi-diagonalizes the leptoquark mass matrices by making
\bs
\label{jgfogh}
\ba
\delta_{1,7} &=& H_1 h,
\label{jcvuifgoi} \\
\left( \begin{array}{c} \delta_{-1,7} \\ \delta_{1,1} \\ \tau_{2,-2}
\end{array} \right) &=& \left( \begin{array}{c} U_1 \\ U_2 \\ U_3
\end{array} \right) u,
\\
\left( \begin{array}{c} \sigma_{0,-2} \\ \delta_{-1,1} \\ \tau_{0,-2}
\end{array} \right) &=& \left( \begin{array}{c} D_1 \\ D_2 \\ D_3
\end{array} \right) d,
\\
\left( \begin{array}{c} \sigma_{0,-8} \\ \tau_{-2,-2}
\end{array} \right) &=& \left( \begin{array}{c} L_1 \\ L_2
\end{array} \right) l,
\ea
\es
where $h$ in Eq.~\eqref{jcvuifgoi} stands for a column matrix
containing the $n_h$ $h$-type scalars;
and analogously for $u$,
$d$,
and $l$ in the other three Eqs.~\eqref{jgfogh}.
The matrices $H_1$,
$U_1$,
$U_2$,
$U_3$,
$D_1$,
$D_2$,
$D_3$,
$L_1$,
and $L_2$ have dimensions $n_{\delta,7} \times n_h$,
$n_{\delta,7} \times n_u$,
$n_{\delta,1} \times n_u$,
$n_{\tau,-2} \times n_u$,
$n_{\sigma,-2} \times n_d$,
$n_{\delta,1} \times n_d$,
$n_{\tau,-2} \times n_d$,
$n_{\sigma,-8} \times n_l$,
and $n_{\tau,-2} \times n_l$,
respectively.
The matrices
\be
\label{unit}
H_1, \quad \left( \begin{array}{c} U_1 \\ U_2 \\ U_3
\end{array} \right), \quad \left( \begin{array}{c} D_1 \\ D_2 \\ D_3
\end{array} \right), \quad \mbox{and} \ \left( \begin{array}{c} L_1 \\ L_2
\end{array} \right)
\ee
are unitary.\footnote{Without loss of generality,
one may chose a basis where $H_1$ is equal to the unit matrix.
We refrain from doing that,
in order to keep the notation as general as possible.}

\paragraph{Mixing matrices:}
We next define the mixing matrices that appear in the charged-current
interactions of the scalars.
They are
\bs
\label{NVQ}
\ba
\N &=& H_1^\dagger U_1,
\\
\V &=& U_2^\dagger D_2 + \sqrt{2}\, U_3^\dagger D_3,
\\
\Q &=& \sqrt{2}\, D_3^\dagger L_2,
\ea
\es
respectively.
The factors $\sqrt{2}$ in Eqs.~\eqref{NVQ} arise
because the charged gauge interactions of the triplets
are $\sqrt{2}$ times stronger than those of the doublets.
The mixing matrices that appear in the neutral-current interactions
of the scalars are
\bs
\label{bars}
\ba
\bar H &=& \Hh - 2 Q_h s_W^2 \times \mathbbm{1}_{n_h},
\\
\bar U &=& \Uu - 2 Q_u s_W^2 \times \mathbbm{1}_{n_u},
\\
\bar D &=& \Dd + 2 Q_d s_W^2 \times \mathbbm{1}_{n_d},
\\
\bar L &=& \Ll + 2 Q_l s_W^2 \times \mathbbm{1}_{n_l},
\ea
\es
where $\mathbbm{1}_m$ denotes the $m \times m$ unit matrix
and the matrices $\Hh$,
$\Uu$,
$\Dd$,
and $\Ll$ are related to the mixing matrices in Eqs.~\eqref{NVQ} through
\bs
\label{vcvufgio}
\ba
\Hh &=& \N \N^\dagger,
\\
\Uu &=& \V \V^\dagger - \N^\dagger \N,
\\
\Dd &=& \V^\dagger \V - \Q \Q^\dagger,
\\
\Ll &=& \Q^\dagger \Q,
\ea
\es
respectively.

\paragraph{Gauge interactions:}
The gauge interactions of the scalars are given by
the following pieces of the Lagrangian:
\bs
\ba
\mathcal{L}_{A S S} &=& - i g s_W A_\theta \left[
  Q_h \sum_{h} \left( h^\ast \partial^\theta h - h \partial^\theta h^\ast \right)
  + Q_u \sum_{u} \left( u^\ast \partial^\theta u - u \partial^\theta u^\ast \right)
  \right. \no & & \left.
  + Q_d \sum_{d} \left( d^\ast \partial^\theta d - d \partial^\theta d^\ast \right)
  + Q_l \sum_{l} \left( l^\ast \partial^\theta l - l \partial^\theta l^\ast \right)
  \right],
\\
\mathcal{L}_{A A S S} &=& g^2 s_W^2 A_\theta A^\theta
\left( Q_h^2\, \sum_{h} h h^\ast + Q_u^2\, \sum_{u} u u^\ast
+ Q_d^2\, \sum_{d} d d^\ast + Q_l^2\, \sum_{l} l l^\ast \right),
\hspace*{3mm} \\
\mathcal{L}_{W S S} &=&  i\, \frac{g}{\sqrt{2}}\, W_\theta^+ \left[
  \sum_{h,u} \N_{h u} \left(
  h^\ast \partial^\theta u - u \partial^\theta h^\ast \right)
  + \sum_{u,d} \V_{u d} \left(
  u^\ast \partial^\theta d - d \partial^\theta u^\ast \right)
  \right. \no & & \left.
  + \sum_{d,l} \Q_{d l} \left(
  d^\ast \partial^\theta l - l \partial^\theta d^\ast \right)
  \right] + \mathrm{H.c.},
\label{Wphiphi}
\\
\mathcal{L}_{W W S S} &=& \frac{g^2}{2}\, W^+_\theta W^{- \theta} \left[
  \sum_{h, h^\prime} \left( \N \N^\dagger \right)_{h h^\prime} h^\ast h^\prime
  + \sum_{u, u^\prime} \left( \N^\dagger \N + \V \V^\dagger \right)_{u u^\prime}
  u^\ast u^\prime
  \right. \no & & \left.
  + \sum_{d, d^\prime} \left( \V^\dagger \V + \Q \Q^\dagger \right)_{d d^\prime}
  d^\ast d^\prime
  + \sum_{l, l^\prime} \left( \Q^\dagger \Q \right)_{l l^\prime} l^\ast l^\prime
  \right],
\label{WWphiphi}
\\
\mathcal{L}_{Z S S} &=& i\, \frac{g}{2c_W}\, Z_\theta \left[
  \sum_{h} \bar H_{h h^\prime}
  \left( h^\ast \partial^\theta h^\prime
  - h^\prime \partial^\theta h^\ast \right)
  + \sum_{u, u^\prime} \bar U_{u u^\prime}
  \left( u^\ast\, \partial^\theta u^\prime
  - u^\prime\, \partial^\theta u^\ast \right)
  \right. \no & & \left.
  -  \sum_{d, d^\prime} \bar D_{d d^\prime}
  \left( d^\ast\, \partial^\theta d^\prime
  - d^\prime\, \partial^\theta d^\ast \right)
  - \sum_{l, l^\prime} \bar L_{l l^\prime}
  \left( l^\ast\, \partial^\theta l^\prime
  - l^\prime\, \partial^\theta l^\ast \right)
  \right],
\label{Zphiphi}
\\
\mathcal{L}_{Z Z S S} &=& \frac{g^2}{4 c_W^2}\, Z_\theta Z^\theta \left[
  \sum_{h, h^\prime} \left( \bar H^2 \right)_{h h^\prime} h^\ast h^\prime
  + \sum_{u, u^\prime} \left( \bar U^2 \right)_{u u^\prime} u^\ast u^\prime
  \right. \no & & \left.
  + \sum_{d, d^\prime} \left( \bar D^2 \right)_{d d^\prime} d^\ast d^\prime
  + \sum_{l, l^\prime} \left( \bar L^2 \right)_{l l^\prime} l^\ast l^\prime
  \right],
\label{ZZphiphi}
\\
\mathcal{L}_{A Z S S} &=& - \frac{g^2 s_W}{c_W}\, A_\theta Z^\theta \left(
Q_h\, \sum_{h, h^\prime} \bar H_{h h^\prime}\, h^\ast h^\prime
+ Q_u\, \sum_{u, u^\prime} \bar U_{u u^\prime}\, u^\ast u^\prime
\right. \no & & \left.
- Q_d\, \sum_{d, d^\prime} \bar D_{d d^\prime}\, d^\ast d^\prime
- Q_l\, \sum_{l, l^\prime} \bar L_{l l^\prime}\, l^\ast l^\prime
\right).
\label{AZphiphi}
\ea
\es
Notice the presence in Eqs.~\eqref{Wphiphi} and~\eqref{WWphiphi}
of the matrices defined in Eqs.~\eqref{NVQ}
and the presence in Eqs.~\eqref{Zphiphi}--\eqref{AZphiphi} of the matrices
defined in Eqs.~\eqref{bars}.
Also notice that,
out of the four matrices in Eq.~\eqref{WWphiphi},
only $\N \N^\dagger = \Hh$ and $\Q^\dagger \Q = \Ll$
coincide with matrices in Eqs.~\eqref{vcvufgio}.

\section{Tools for the computation}

\paragraph{PV functions:}
The Passarino--Veltman (PV) functions~\cite{pv}
$B_{00} \left( q^2, m_1^2, m_2^2 \right)$ and $A_0 \left( m_1^2 \right)$
are defined by\footnote{We use the definitions of Ref.~\cite{looptools}
for the PV functions.}
\bs
\label{kgfog0}
\ba
\mu^\epsilon \int \frac{\mathrm{d}^{4 - \epsilon} k}{\left( 2 \pi
  \right)^{4 - \epsilon}}\
\, k^\theta k^\psi\, \frac{1}{k^2 - m_1^2}\
\frac{1}{\left( k + q \right)^2 - m_2^2}
&=& \frac{i}{16 \pi^2}\, \left[ g^{\theta \psi}\,
  B_{00} \left( q^2, m_1^2, m_2^2 \right)
  \right. \no & & \left.
  + q^\theta q^\psi\, B_{11} \left( q^2, m_1^2, m_2^2 \right) \right],
\hspace*{7mm}
\label{b00}
\\
\mu^\epsilon \int \frac{\mathrm{d}^{4 - \epsilon} k}{\left( 2 \pi
  \right)^{4 - \epsilon}}\
\frac{1}{k^2 - m_1^2}\
&=& \frac{i}{16 \pi^2}\ A_0 \left( m_1^2 \right),
\ea
\es
respectively,
where $\mu$ is an arbitrary quantity with mass dimension.
The quantities $q^2$,
$m_1^2$,
and $m_2^2$ are assumed to be non-negative.
The PV function $B_{11} \left( q^2, m_1^2, m_2^2 \right)$ in Eq.~\eqref{b00}
is not needed in this paper;
on the other hand,
we need
\bs
\ba
\label{prime}
B_{00}^\prime \left( q^2, m_1^2, m_2^2 \right) &\equiv&
\frac{\partial B_{00} \left( q^2, m_1^2, m_2^2 \right)}{\partial q^2},
\\
\bar B_{00} \left( q^2, m_1^2, m_2^2 \right) &\equiv&
\frac{B_{00} \left( q^2, m_1^2, m_2^2 \right)
  - B_{00} \left( 0, m_1^2, m_2^2 \right)}{q^2}.
\label{bar}
\ea
\es
The PV functions may be numerically evaluated by using softwares like
{\tt LoopTools}~\cite{looptools}
and~{\tt COLLIER}~\cite{collier}.\footnote{For the present purposes,
in the results given by those codes one should take only the \emph{real},
\textit{viz.}\ dispersive,
parts of the PV functions,
while dropping the absorptive parts,
which have no relevance for the oblique parameters.}
They may also be evaluated through analytic formulas
given in Ref.~\cite{romao2}.

\begin{figure}[h!]
  \centering
    \includegraphics[width=0.7\textwidth]{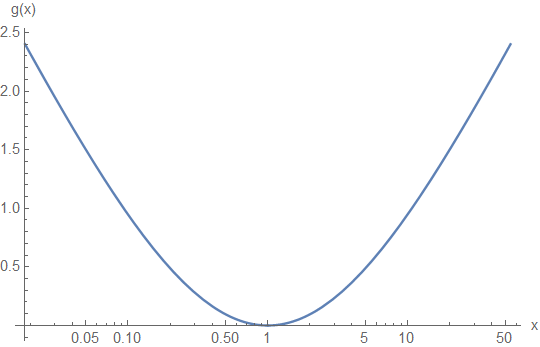}
    \caption{Plot of the function $g \left( x \right)$.
      The scale in the horizontal axis is logarithmic.}
  \label{fig:g}
\end{figure}

\paragraph{The function $g$:}
One has~\cite{romao2}
\be
\label{divv}
B^\prime_{00} \left( 0, m_1^2, m_2^2 \right) = - \frac{\mathrm{div}}{12}
+ \frac{1}{24} \left[ \ln{\frac{m_1^2}{\mu^2}} + \ln{\frac{m_2^2}{\mu^2}}
  + g \left( \frac{m_1^2}{m_2^2} \right) \right],
\ee
where $\mathrm{div}$ is a divergent quantity
which is defined in Eq.~\eqref{div} and
cancels out in the final results,
and
\be
\label{functiong}
g \left( x \right) = \left\{ \begin{array}{lcl}
  {\displaystyle \frac{x^3 - 3 x^2 - 3 x + 1}{\left( x - 1 \right)^3}\, \ln{x}
  - \frac{5 x^2 - 22 x + 5}{3 \left( x - 1 \right)^2}} 
  & \Leftarrow & x \neq 1,
  \\*[2mm]
  0 & \Leftarrow & x = 1
\end{array} \right.
\ee
is a function that obeys $g \left( 1 \right) = 0$
and $g \left( x \right) = g \left( 1/x \right)$.
This function is depicted in Fig.~\ref{fig:g}.

\paragraph{The function $\theta_+$:}
From the formulas in Ref.~\cite{romao2} one gathers that
\be
\label{relation}
B_{00} \left( 0, m_1^2, m_2^2 \right) =
\frac{A_0 \left( m_1^2 \right) + A_0 \left( m_2^2 \right)}{4}
+ \frac{\theta_+ \left( m_1^2, m_2^2 \right)}{8},
\ee
where~\cite{silvalavoura}
\be
\label{theta+}
\theta_+ \left( m_1^2, m_2^2 \right) \equiv \left\{ \begin{array}{lcl}
  \displaystyle{m_1^2 + m_2^2 - \frac{2 m_1^2 m_2^2}{m_1^2 - m_2^2}\,
    \ln{\frac{m_1^2}{m_2^2}}} & \Leftarrow & m_1^2 \neq m_2^2,
  \\*[2mm]
  0 & \Leftarrow & m_1^2 = m_2^2.
\end{array} \right.
\ee

\paragraph{The function $\rho$:}
We define the function $\rho \left( x, y \right)$---where $x$ and $y$
are positive---through
\be
\label{functionrho}
\rho \left( x, y \right)
\equiv 2 \left[ B_{00}^\prime \left( q^2, q^2 x, q^2 y \right)
- \bar B_{00} \left( q^2, q^2 x, q^2 y \right) \right],
\ee
which is $q^2$-independent.
Using the analytic formulas in ref.~\cite{romao2} one finds that
\be
\label{681}
\rho \left( x, y \right) = \left\{ \begin{array}{lcl}
\displaystyle{\frac{1}{6} - \frac{3 \left( x + y \right)}{4}
+ \frac{\left( x - y \right)^2}{2}
+ \left[ \frac{\left( y - x \right)^3}{4}
  + \frac{x^2 + y^2}{4 \left( y - x \right)} \right.} & &
\\*[2mm]
\displaystyle{\left. + \frac{x^2 - y^2}{2} \right] \ln{\frac{x}{y}}
+ \frac{\left( x - y \right)^2 - x - y}{4}\
f \left( x, y \right)} & \Leftarrow & x \neq y,
\\*[4mm]
\displaystyle{\frac{1}{6} - 2 x - \frac{x}{2}\ f \left( x, x \right)}
& \Leftarrow & x = y.
\end{array}
\right.
\ee
In Eq.~\eqref{681},
\be
f \left( x, y \right) = \left\{
\begin{array}{lcl}
  \displaystyle{\sqrt{\Delta \left( x, y \right)}\
    \ln{\frac{x + y - 1 + \sqrt{\Delta \left( x, y \right)}}{x + y
        - 1 - \sqrt{\Delta \left( x, y \right)}}}}
    & \Leftarrow & \Delta \left( x, y \right) \ge 0,
  \\*[4mm]
  \displaystyle{- 2 \sqrt{- \Delta \left( x, y \right)} \left[
    \arctan{\frac{x - y + 1}{\sqrt{- \Delta \left( x, y \right)}}}
    + \left( x \leftrightarrow y \right) \right]}
    & \Leftarrow & \Delta \left( x, y \right) < 0,
\end{array}
\right.
\ee
where
\be
\Delta \left( x, y \right)
= 1 - 2 \left( x + y \right) + \left( x - y \right)^2.
\ee

\begin{figure}[h!]
\centering
   \begin{subfigure}{0.49\linewidth} \centering
     \includegraphics[scale=0.27]{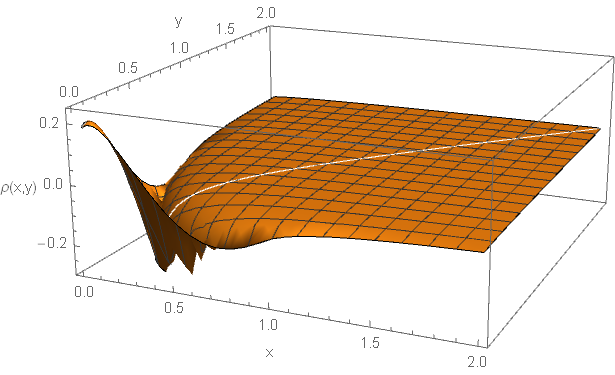}
   \end{subfigure}
   \begin{subfigure}{0.49\linewidth} \centering
     \includegraphics[scale=0.27]{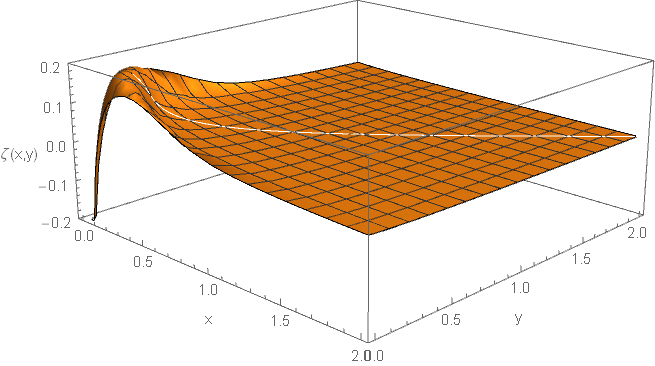}
   \end{subfigure}
   \caption{Plots of the functions $\rho \left( x , y \right)$
     (left panel) and $\zeta \left( x, y \right)$
     (right panel).} 
\label{fig:rhozeta}
\end{figure}

\paragraph{The function $\zeta$:}
The function $\zeta \left( x, y \right)$---where $x$ and $y$
are positive---is defined through
\be
\label{functionzeta}
\zeta \left( x, y \right)
\equiv 2 \left[ B_{00}^\prime \left( 0, q^2 x, q^2 y \right)
- \bar B_{00} \left( q^2, q^2 x, q^2 y \right) \right],
\ee
which is $q^2$-independent.
Using the analytic formulas in ref.~\cite{romao2} one finds that
\be
\label{zeta1}
\zeta \left( x, y \right) = \left\{ \begin{array}{lcl}
\displaystyle{\frac{11}{36} - \frac{5 \left( x + y \right)}{12}
+ \frac{x y}{3 \left( x - y \right)^2}
+ \frac{\left( x - y \right)^2}{6}
+ \left[
  \frac{x^2 - y^2}{4}
  + \frac{\left( y - x \right)^3}{12} \right.} & &
\\*[2mm]
\displaystyle{\left. + \frac{x^2 + y^2}{4 \left( y - x \right)}
  + \frac{x + y}{12 \left( x - y \right)}
  + \frac{x y \left( x + y \right)}{6 \left( y - x \right)^3} 
  \right] \ln{\frac{x}{y}}} & &
\\*[2mm]
\displaystyle{+ \frac{\Delta \left( x, y \right)}{12}\ f \left( x, y \right)}
& \Leftarrow & x \neq y,
\\*[3mm]
\displaystyle{\frac{4}{9} - \frac{4 x}{3}
  + \frac{\Delta \left( x, x \right)}{12}\ f \left( x, x \right)}
& \Leftarrow & x = y.
\end{array} \right.
\ee
The functions $\rho \left( x, y \right)$ and $\zeta \left( x, y \right)$
are illustrated in Fig.~\ref{fig:rhozeta}.
They are both very small when $x \gtrsim 1$ and $y \gtrsim 1$.

\paragraph{Computation of the diagrams:}
Suppose the vertices $V_\mu S_1 S_2$ and $V_\mu V^\prime_\nu S_1 S_2$
have Feynman rules
\vspace{5mm}
\begin{subequations}
\allowdisplaybreaks
\begin{align}
\parbox{35mm}{
    \begin{fmfgraph*}(75,75)
    \fmfleft{i1,i2}
    \fmfright{o1}
    \fmf{dashes}{i1,w1}
    \fmf{dashes}{i2,w1}
    \fmf{photon}{w1,o1}
    \fmflabel{$S_2$}{i1}
    \fmflabel{$S_1$}{i2}
    \fmflabel{$V_\mu$}{o1}
    \marrow{a}{left}{lft}{$k$}{i1,w1}{(1/4, 4/5)}
    \marrow{b}{left}{lft}{$p$}{i2,w1}{(1/4, 4/5)}
\end{fmfgraph*}}  &= i X \left(k - p\right)_\mu,  \\[40pt]
\parbox{35mm}{
    \begin{fmfgraph*}(75,75)
    \fmfleft{i1,i2}
    \fmfright{o1,o2}
    \fmf{dashes}{i1,w1}
    \fmf{dashes}{i2,w1}
    \fmf{photon}{w1,o1}
    \fmf{photon}{w1,o2}
    \fmflabel{$S_2$}{i1}
    \fmflabel{$S_1$}{i2}
    \fmflabel{$V_\mu$}{o2}
    \fmflabel{$V_\nu^\prime$}{o1}
\end{fmfgraph*}}  &= i Y g_{\mu \nu},
\end{align}
\end{subequations}

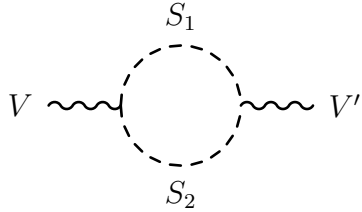
\begin{figure}[h!]
  \centering
    \begin{fmfgraph*}(100,100)
    \fmfleft{i1}
    \fmfright{o1}
    \fmf{boson,label.side=left}{i1,w1}
    \fmf{boson,label.side=left}{w2,o1}
    \fmf{dashes,right,tension=.3,label=$S_2$}{w1,w2}
    \fmf{dashes,left,tension=.3,label=$S_1$}{w1,w2}
    \fmflabel{$V$}{i1}
    \fmflabel{$V^\prime$}{o1}
    \end{fmfgraph*}
  \caption{One type of diagram for vacuum polarization.
$V$ and $V^\prime$ are gauge bosons,
$S_1$ is a scalar with mass $m_1$,
and $S_2$ is a scalar with mass $m_2$.}
  \label{fig1}
\end{figure}

\begin{figure}[h!]
  \centering
    \begin{fmfgraph*}(100,100)
    \fmfleft{i1}
    \fmfright{o1}
    \fmf{boson,label.side=left}{i1,w1}
    \fmf{boson,label.side=left}{w1,o1}
    \fmf{dashes,label=$S_1$}{w1,w1}
    \fmflabel{$V$}{i1}
    \fmflabel{$V^\prime$}{o1}
    \end{fmfgraph*}
    \vspace{-12mm}
  \caption{Another type of diagram for vacuum polarization.
$V$ and $V^\prime$ are gauge bosons
and $S_1$ is a scalar with mass $m_1$.}
  \label{fig2}
\end{figure}
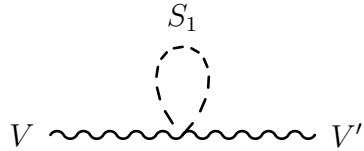

\vspace{7mm}

\noindent respectively,
where $V$ and $V^\prime$ are gauge bosons
(\textit{v.g.}\ either $A$,
$Z$,
$W^+$,
or $W^-$)
and $S_1$ and $S_2$ are scalars.
Using those Feynman rules we have
\be
\label{as1s2}
A^{V V^\prime S_1 S_2} \left( q^2 \right) =
\frac{X_1 X_2}{4 \pi^2}\, B_{00} \left( q^2, m_1^2, m_2^2 \right)
\ee
for diagrams of the form in Fig.~\ref{fig1},
and
\be
\label{as1}
A^{V V^\prime S_1} \left( q^2 \right)
= - \frac{Y}{16 \pi^2}\, A_0 \left( m_1^2 \right)
\ee
for diagrams of the form in Fig.~\ref{fig2}.
In Eqs.~\eqref{as1s2} and~\eqref{as1},
$q$ is the four-momentum of the gauge bosons.

\section{Results for the oblique parameters}

\paragraph{$T$:} For the oblique parameter $T$ we have
\bs
\label{t1}
\ba
T &=& \frac{1}{8 \pi s_W^2 m_W^2} \left\{
  4 \left[ \sum_{h,u} \left| \N_{h u} \right|^2 B_{00}(0, m_h^2, m_u^2)
    + \sum_{u,d} \left| \V_{u d} \right|^2 B_{00}(0, m_u^2, m_d^2)
    \right. \right. \no & & \left.
    + \sum_{d,l} \left| \Q_{d l} \right|^2 B_{00}(0, m_d^2, m_l^2) \right]
  \label{1T} \\ & &
  - \sum_h \left( \N \N^\dagger \right)_{hh}\, A_0 \left( m_h^2 \right)
  - \sum_u \left( \N^\dagger \N + \V \V^\dagger \right)_{uu}\,
  A_0 \left( m_u^2 \right)
  \no & &
  - \sum_d \left( \V^\dagger \V + \Q \Q^\dagger \right)_{dd}\,
  A_0 \left( m_d^2 \right)
  - \sum_l \left( \Q^\dagger \Q \right)_{ll}\, A_0 \left( m_l^2 \right)
  \label{T2} \\ & &
  - 2 \left[\sum_{h, h^\prime} \left| \bar H_{h h^\prime} \right|^2
  B_{00} \left( 0, m_h^2, m_{h^\prime}^2 \right)
  + \sum_{u, u^\prime} \left| \bar U_{u u^\prime} \right|^2
  B_{00} \left( 0, m_u^2, m_{u^\prime}^2 \right) 
  \right. \no & & \left.
  + \sum_{d, d^\prime} \left| \bar D_{d d^\prime} \right|^2
  B_{00} \left( 0, m_d^2, m_{d^\prime}^2 \right) 
  + \sum_{l, l^\prime} \left| \bar L_{l l^\prime} \right|^2
  B_{00} \left( 0, m_l^2, m_{l^\prime}^2 \right) \right]
  \label{T3} \\ & &
  + \sum_h \left( \bar H^2 \right)_{hh}\, A_0 \left( m_h^2 \right)
  + \sum_u \left( \bar U^2 \right)_{uu}\, A_0 \left( m_u^2 \right)
  \no & & \left.
  + \sum_d \left( \bar D^2 \right)_{dd}\, A_0 \left( m_d^2 \right)
  + \sum_l \left( \bar L^2 \right)_{ll}\, A_0 \left( m_l^2 \right)
  \right\},
  \label{T4}
\ea
\es
where lines~\eqref{1T} and~\eqref{T2} originate in $A_{WW} \left( 0 \right)$
while lines~\eqref{T3} and~\eqref{T4} originate in $A_{ZZ} \left( 0 \right)$.
Utilizing Eq.~\eqref{relation} on Eq.~\eqref{t1},
we obtain
\ba
T &=& \frac{1}{16 \pi s_W^2 m_W^2} \left[
\sum_{h,u} \left| \N_{hu} \right|^2 \theta_+ \left( m_h^2, m_u^2 \right)
\right. \no & &
+ \sum_{u,d} \left| \V_{ud} \right|^2 \theta_+ \left( m_u^2, m_d^2 \right)
+ \sum_{d,l} \left| \Q_{dl} \right|^2 \theta_+ \left( m_d^2, m_l^2 \right)
\no & &
- \sum_{h < h^\prime} \left| \Hh_{h h^\prime} \right|^2
\theta_+ \left( m_h^2, m_{h^\prime}^2 \right)
- \sum_{u < u^\prime} \left| \Uu_{u u^\prime} \right|^2
\theta_+ \left( m_u^2, m_{u^\prime}^2 \right)
\no & & \left.
- \sum_{d < d^\prime} \left| \Dd_{d d^\prime} \right|^2
\theta_+ \left( m_d^2, m_{d^\prime}^2 \right)
- \sum_{l < l^\prime} \left| \Ll_{l l^\prime} \right|^2
\theta_+ \left( m_l^2, m_{l^\prime}^2 \right)
\right],
\label{t2}
\ea
The function $\theta_+ \left( m_1^2, m_2^2 \right)$
is given in Eq.~\eqref{theta+}.
It is a finite function;
thus,
in Eq.~\eqref{t2} there are no divergences,
contrary to what happens in Eq.~\eqref{t1},
wherein the functions $B_{00} \left( 0, m_1^2, m_2^2 \right)$
and $A_0 \left( m_3^2 \right)$ are divergent.
The function $\theta_+ \left( m_1^2, m_2^2 \right)$
is moreover positive definite;
thus,
the two first lines of Eq.~\eqref{t2} constitute a positive contribution to $T$,
which always exists in a model with leptoquarks,
while the two last lines of that equation are a negative contribution,
which only exists if at least one of the matrices $\Hh$,
$\Uu$,
$\Dd$,
and $\Ll$ has nonzero off-diagonal matrix elements.

\paragraph{$V$ and $W$:}
For the oblique parameters $V$ and $W$ we obtain the results
\bs
\label{33}
\ba
V &=& \frac{1}{8 \pi c_W^2 s_W^2} \left[
\sum_{h,h^\prime} \left| \bar H_{h h^\prime} \right|^2
\rho \left( \frac{m_h^2}{m_Z^2}, \frac{m_{h^\prime}^2}{m_Z^2} \right)
+ \sum_{u,u^\prime} \left| \bar U_{u u^\prime} \right|^2
\rho \left( \frac{m_u^2}{m_Z^2}, \frac{m_{u^\prime}^2}{m_Z^2} \right)
\right. \no & & \left.
+ \sum_{d,d^\prime} \left| \bar D_{d d^\prime} \right|^2
\rho \left( \frac{m_d^2}{m_Z^2}, \frac{m_{d^\prime}^2}{m_Z^2} \right)
+ \sum_{l,l^\prime} \left| \bar L_{l l^\prime} \right|^2
\rho \left(\frac{m_l^2}{m_Z^2}, \frac{m_{l^\prime}^2}{m_Z^2} \right)
\right],
\\
W &=& \frac{1}{4 \pi s_W^2} \left[
  \sum_{h, u} \left| \N_{hu} \right|^2
  \rho \left( \frac{m_h^2}{m_W^2}, \frac{m_u^2}{m_W^2} \right)
  + \sum_{u, d} \left| \V_{ud} \right|^2
  \rho \left( \frac{m_u^2}{m_W^2}, \frac{m_d^2}{m_W^2} \right)
  \right. \no & & \left.
  + \sum_{d, l} \left| \Q_{dl} \right|^2
  \rho \left( \frac{m_d^2}{m_W^2}, \frac{m_l^2}{m_W^2} \right)
\right],
\ea
\es
where the function $\rho$ is given in Eq.~\eqref{681}.

\paragraph{$X$:}

For the oblique parameter $X$ one has
\ba
X &=& - \frac{1}{4 \pi} \left[
  Q_h \sum_h \bar H_{hh}\
  \zeta \left( \frac{m_h^2}{m_Z^2}, \frac{m_h^2}{m_Z^2} \right)
  + Q_u \sum_u \bar U_{uu}\
  \zeta \left( \frac{m_u^2}{m_Z^2}, \frac{m_u^2}{m_Z^2} \right)
  \right. \no & & \left.
  - Q_d \sum_d \bar D_{dd}\
  \zeta \left( \frac{m_d^2}{m_Z^2}, \frac{m_d^2}{m_Z^2} \right)
  - Q_l \sum_l \bar L_{ll}\
  \zeta \left( \frac{m_l^2}{m_Z^2}, \frac{m_l^2}{m_Z^2} \right)
  \right],
\label{XXXXX}
\ea
where the function $\zeta$ is given in Eq.~\eqref{zeta1}.

\paragraph{$S$ and $U$:}

Following the decomposition in Eqs.~\eqref{dec1}--\eqref{dec3},
for the oblique parameters $S$ and $U$ we have
\bs
\label{tttt1}
\ba
S^{\prime \prime} &=& - \frac{1}{2 \pi} \left[
  \sum_{h, h^\prime} \left| \bar H_{h h^\prime} \right|^2
  \zeta \left( \frac{m_h^2}{m_Z^2}, \frac{m_{h^\prime}^2}{m_Z^2} \right)
  + \sum_{u, u^\prime} \left| \bar H_{u u^\prime} \right|^2
  \zeta \left( \frac{m_u^2}{m_Z^2}, \frac{m_{u^\prime}^2}{m_Z^2} \right)
  \right. \no & & \left.
  + \sum_{d, d^\prime} \left| \bar H_{d d^\prime} \right|^2
  \zeta \left( \frac{m_d^2}{m_Z^2}, \frac{m_{d^\prime}^2}{m_Z^2} \right)
  + \sum_{l, l^\prime} \left| \bar H_{l l^\prime} \right|^2
  \zeta \left( \frac{m_l^2}{m_Z^2}, \frac{m_{l^\prime}^2}{m_Z^2} \right)
  \right],
\\
U^{\prime \prime} &=& - S^{\prime \prime} - \frac{1}{\pi} \left[
  \sum_{h, u} \left| \N_{hu} \right|^2
  \zeta \left( \frac{m_h^2}{m_W^2}, \frac{m_u^2}{m_W^2} \right)
  + \sum_{u, d} \left| \V_{ud} \right|^2
  \zeta \left( \frac{m_u^2}{m_W^2}, \frac{m_d^2}{m_W^2} \right)
  \right. \no & & \left.
  + \sum_{d, l} \left| \Q_{dl} \right|^2
  \zeta \left( \frac{m_d^2}{m_W^2}, \frac{m_l^2}{m_W^2} \right)
  \right],
\ea
\es
where the function $\zeta$ is given in Eq.~\eqref{zeta1};
and
\bs
\label{Sresult}
\ba
S^\prime &=& \frac{1}{\pi} \left\{
\sum_{h,h^\prime} \left| \Hh_{h h^\prime} \right|^2
B_{00}^\prime \left( 0, m_h^2, m_{h^\prime}^2 \right)
+ \sum_{u,u^\prime} \left| \Uu_{uu^\prime} \right|^2
B_{00}^\prime \left( 0, m_u^2, m_{u^\prime}^2 \right)
\right. \no & &
+ \sum_{d,d^\prime} \left| \Dd_{dd^\prime} \right|^2
B_{00}^\prime \left( 0, m_d^2, m_{d^\prime}^2 \right)
+ \sum_{l,l^\prime} \left| \Ll_{ll^\prime} \right|^2
B_{00}^\prime \left( 0, m_l^2, m_{l^\prime}^2 \right)
\no & &
- 2 \left[
  Q_h\, \sum_{h} \Hh_{h h}\, B_{00}^\prime \left( 0, m_h^2, m_h^2 \right)
  + Q_u\, \sum_{u} \Uu_{uu}\, B_{00}^\prime \left( 0, m_u^2, m_u^2 \right)
  \right]
\no & & \left.
+ 2 \left[
  Q_d\, \sum_d \Dd_{dd}\, B_{00}^\prime \left( 0, m_d^2, m_d^2 \right)
  + Q_l\, \sum_l \Ll_{ll}\, B_{00}^\prime \left( 0, m_l^2, m_l^2 \right)
  \right]
\right\},
\label{s3}
\\
\label{S+Uresult}
%\ba
U^\prime &=& - S^\prime + \frac{2}{\pi} \left[
  \sum_{h,u} \left| \N_{hu} \right|^2
  B_{00}^\prime \left( 0, m_h^2, m_u^2 \right)
  + \sum_{u,d} \left| \V_{ud} \right|^2
  B_{00}^\prime \left( 0, m_u^2, m_d^2 \right)
  \right. \no & &
  + \sum_{d,l} \left| \Q_{dl} \right|^2
  B_{00}^\prime \left( 0, m_d^2, m_l^2 \right)
  \no & &
  - Q_h\, \sum_{h} \Hh_{hh}\, B_{00}^\prime \left( 0, m_h^2, m_h^2 \right)
  - Q_u\, \sum_{u} \Uu_{uu}\, B_{00}^\prime \left( 0, m_u^2, m_u^2 \right)
  \no & & \left.
  + Q_d\, \sum_d \Dd_{dd}\, B_{00}^\prime \left( 0, m_d^2, m_d^2 \right)
  + Q_l\, \sum_l \Ll_{ll}\, B_{00}^\prime \left( 0, m_l^2, m_l^2 \right)
  \right].
\label{u4}
\ea
\es
Utilizing Eq.~\eqref{divv},
and the cancellation of the divergences proved in Appendix~C,
one simplifies Eqs.~\eqref{Sresult} to
\bs
%\allowdisplaybreaks
\label{t5}
\ba
S^\prime &=&  \frac{1}{12 \pi} \left\{
\sum_{h < h^\prime} \left| \Hh_{h h^\prime} \right|^2
  g \left( \frac{m_h^2}{m_{h^\prime}^2} \right)
+ \sum_{u < u^\prime} \left| \Uu_{uu^\prime} \right|^2
  g \left(\frac{m_u^2}{m_{u^\prime}^2} \right) 
\right.  \label{A} \\ & &
+ \sum_{d < d^\prime} \left| \Dd_{dd^\prime} \right|^2
 g \left( \frac{m_d^2}{m_{d^\prime}^2} \right) 
+ \sum_{l < l^\prime} \left|  \Ll_{ll^\prime} \right|^2
 g \left( \frac{m_l^2}{m_{l^\prime}^2} \right) 
 \label{B} \\ & & 
 + \sum_h \left[ \left( \Hh^2 \right)_{hh} - 2 Q_h \Hh_{hh}\right]
 \ln{\frac{m_h^2}{\mu^2}}
 + \sum_u \left[ \left( \Uu^2 \right)_{uu} - 2 Q_u \Uu_{uu}\right]
 \ln{\frac{m_u^2}{\mu^2}}
 \label{C} \\ & & \left.
 + \sum_d \left[ \left( \Dd^2 \right)_{dd} + 2 Q_d \Dd_{dd} \right]
 \ln{\frac{m_d^2}{\mu^2}}
 + \sum_l \left[ \left( \Ll^2 \right)_{ll} + 2 Q_l \Ll_{ll} \right]
 \ln{\frac{m_l^2}{\mu^2}}
\right\},
\label{D} \\
U^\prime &=& - S^\prime + \frac{1}{12 \pi} \left\{
\sum_{h,u} \left| \N_{hu} \right|^2 g \left( \frac{m_h^2}{m_u^2} \right)
+ \sum_{u,d} \left| \V_{ud} \right|^2 g \left( \frac{m_u^2}{m_d^2} \right)
+ \sum_{d,l} \left| \Q_{dl} \right|^2 g \left( \frac{m_d^2}{m_l^2} \right)
\right. \hspace*{7mm} \label{E} \\ & &
+ \left( 1 - 2 Q_h \right) \sum_h \Hh_{hh} \ln{\frac{m_h^2}{\mu^2}}
+ \sum_u \left[ \left( \N^\dagger \N + \V \V^\dagger \right)_{uu}
  - 2 Q_u \Uu_{uu} \right] \ln{\frac{m_u^2}{\mu^2}}
\label{F} \\ & & \left.
+ \sum_d \left[ \left( \V^\dagger \V + \Q \Q^\dagger \right)_{dd}
  + 2 Q_d \Dd_{dd} \right] \ln{\frac{m_d^2}{\mu^2}}
+ \left( 1 + 2 Q_l \right) \sum_l \Ll_{ll} \ln{\frac{m_l^2}{\mu^2}}
\right\}. \label{G}
\ea
\es
The function $g$ in lines~\eqref{A}, \eqref{B}, and~\eqref{E}
is given in Eq.~\eqref{functiong}.
In lines~\eqref{C},
\eqref{D},
\eqref{F},
and~\eqref{G} $\mu$ is the arbitrary mass introduced in Eqs.~\eqref{kgfog0};
$S^\prime$ and $U^\prime$ do not depend on the value of $\mu$,
just as they do not depend on the divergent quantity of Eq.~\eqref{div}.

\section{Models with few parameters}

We present in this section three examples of leptoquark models,
each of them with two leptoquark multiplets and one mixing angle.
For the sake of simplicity,
for each model we only write down the expressions
for the three original oblique parameters $T$,
$S^\prime$,
and $U^\prime$.
Note that we multiply the original Eqs.~\eqref{t2} and~\eqref{t5}
by a factor 3 because of the three possible colours of each leptoquark.

\subsection{$\left( n_{\sigma,-2}, n_{\sigma,-8}, n_{\delta,7}, n_{\delta,1},
  n_{\tau,-2} \right) = \left( 1, 0, 0, 1, 0 \right)$}

There is one $u$-type scalar with mass $m_u$
and there are two $d$-type scalars with masses $m_{d_1}$ and $m_{d_2}$.
The matrices $U_2 = 1$,
$D_1 = \left( c_\theta, \ -s_\theta \right)$,
and $D_2 = \left( s_\theta, \ c_\theta \right)$,
where $\theta$ is a mixing angle.
Therefore,
the matrices $\N$, $\Q$, $\Hh$, and $\Ll$ do not exist,
$\V = \left( s_\theta, \ c_\theta \right)$,
$\Uu = 1$,
and $\Dd = \begin{pmatrix}
    s_\theta^2 & c_\theta s_\theta \\
    c_\theta s_\theta & c_\theta^2
\end{pmatrix}$.
So,
\bs
\allowdisplaybreaks
\ba
T &=& \frac{3}{16 \pi s_W^2 m_W^2} \left[
  s_\theta^2\ \theta_+ \left( m_u^2, m_{d_1}^2 \right)
  + c_\theta^2\ \theta_+ \left( m_u^2, m_{d_2}^2 \right)
  - c_\theta^2 s_\theta^2\ \theta_+ \left( m_{d_1}^2, m_{d_2}^2 \right)
  \right],
\\
S^\prime &=&  \frac{1}{12 \pi} \left[
  3\, c_\theta^2 s_\theta^2\ g \left( \frac{m_{d_1}^2}{m_{d_2}^2} \right)
  + s_\theta^2\, \ln{\frac{m_{d_1}^2}{m_u^2}}
  + c_\theta^2\, \ln{\frac{m_{d_2}^2}{m_u^2}}
  \right],
\\
U^\prime &=& \frac{1}{4 \pi} \left[
  s_\theta^2\, g \left( \frac{m_u^2}{m_{d_1}^2} \right)
  + c_\theta^2\, g \left( \frac{m_u^2}{m_{d_2}^2} \right)
  - c_\theta^2 s_\theta^2\ g \left( \frac{m_{d_1}^2}{m_{d_2}^2} \right)
  \right].
\ea
\es

\subsection{$\left( n_{\sigma,-2}, n_{\sigma,-8}, n_{\delta,7}, n_{\delta,1},
  n_{\tau,-2} \right) = \left( 0, 0, 1, 1, 0 \right)$}

The is one $h$-type leptoquark with mass $m_h$,
one $d$-type leptoquark with mass $m_d$,
and two up-type leptoquarks with masses $m_{u1}$ and $m_{u2}$.
The matrices $U_3$,
$D_1$,
$D_3$,
$L_1$,
and $L_2$ do not exist;
the matrices $H_1 = D_2 = 1$ are just numbers.
We have $\N = U_1 = \left( c_\theta, \ - s_\theta \right)$
and $\V^\dagger = U_2 = \left( s_\theta, \ c_\theta \right)$;
the matrices $\Q$ and $\Ll$ do not exist,
$\Hh = \Dd = 1$,
and $\Uu = \left( \begin{array}{cc}
  s_\theta^2 - c_\theta^2 & 2 s_\theta c_\theta \\ 2 s_\theta c_\theta &
  c_\theta^2 - s_\theta^2
\end{array} \right)$.
So,
\bs
\ba
T &=& \frac{3}{16 \pi s_W^2 m_W^2} \left\{
c_\theta^2 \left[ \theta_+ \left( m_h^2, m_{u1}^2 \right)
  + \theta_+ \left( m_{u2}^2, m_d^2 \right) \right]
+ s_\theta^2 \left[ \theta_+ \left( m_h^2, m_{u2}^2 \right)
  + \theta_+ \left( m_{u1}^2, m_d^2 \right) \right]
\right. \no & & \left.
- 4 c_\theta^2 s_\theta^2\, \theta_+ \left( m_{u1}^2, m_{u2}^2 \right)
\right\},
\\
S^\prime &=& \frac{c_\theta^2 s_\theta^2}{\pi}\,
g \left( \frac{m_{u1}^2}{m_{u2}^2} \right)
+ \frac{1}{12 \pi} \left( 7 \ln{\frac{m_{u1}^2}{m_h^2}}
+ \ln{\frac{m_d^2}{m_{u2}^2}}
+ 8 s_\theta^2\, \ln{\frac{m_{u2}^2}{m_{u1}^2}} \right),
\\
U^\prime &=& \frac{1}{4 \pi} \left\{
c_\theta^2 \left[ g \left( \frac{m_h^2}{m_{u1}^2} \right)
  + g \left( \frac{m_{u2}^2}{m_d^2} \right) \right]
+ s_\theta^2 \left[ g \left( \frac{m_h^2}{m_{u2}^2} \right)
  + g \left( \frac{m_{u1}^2}{m_d^2} \right) \right]
%\right. \no & & \left.
- 4 c_\theta^2 s_\theta^2\, g \left( \frac{m_{u1}^2}{m_{u2}^2} \right) \right\}.
\no & &
\ea
\es

\subsection{$\left( n_{\sigma,-2}, n_{\sigma,-8}, n_{\delta,7}, n_{\delta,1},
  n_{\tau,-2} \right) = \left( 0, 0, 1, 0, 1 \right)$}

There are one $h$-type leptoquark with mass $m_h$,
two up-type leptoquarks with masses $m_{u1}$ and $m_{u2}$,
respectively,
one $d$-type leptoquark with mass $m_d$,
and one $l$-type leptoquark with mass $m_l$.
The matrices $U_2$,
$D_1$,
$D_2$,
and $L_1$ do not exist;
the matrices $H_1 = D_3 = L_2 = 1$ are just numbers,
while $\N = U_1 = \left( c_\theta, \ - s_\theta \right)$
and
$\V^\dagger \left/ \sqrt{2} \right. = U_3 = \left( s_\theta, \ c_\theta \right)$.
The matrix $\Q = \sqrt{2}$ is a number.
Furthermore,
$\Hh = 1$,
$\Dd = 0$,
$\Ll = 2$,
and $\Uu = \left( \begin{array}{cc}
  3 s_\theta^2 - 1 & 3 s_\theta c_\theta \\ 3 s_\theta c_\theta & 3 c_\theta^2 - 1
\end{array} \right)$.
So,
\bs
\ba
T &=& \frac{3}{16 \pi s_W^2 m_W^2}
\left\{ c_\theta^2 \left[ \theta_+ \left( m_h^2, m_{u1}^2 \right)
  + 2\, \theta_+ \left( m_{u2}^2, m_d^2 \right) \right]
\right. \nonumber \\*[1mm] & &
+ s_\theta^2 \left[ \theta_+ \left( m_h^2, m_{u2}^2 \right)
  + 2\, \theta_+ \left( m_{u1}^2, m_d^2 \right) \right]
\nonumber \\*[1mm] & & \left.
+ 2\, \theta_+ \left( m_d^2, m_l^2 \right)
- 9 c_\theta^2 s_\theta^2\, \theta_+ \left( m_{u1}^2, m_{u2}^2 \right)
\right\},
\\*[1mm]
S^\prime &=& \frac{9 c_\theta^2 s_\theta^2}{4 \pi}\,
g \left( \frac{m_{u1}^2}{m_{u2}^2} \right)
+ \frac{1}{12 \pi} \left( 7 \ln{\frac{m_{u1}^2}{m_h^2}}
+ 4 \ln{\frac{m_{u2}^2}{m_l^2}}
+ 3 s_\theta^2\, \ln{\frac{m_{u2}^2}{m_{u1}^2}} \right),
\\*[1mm]
U^\prime &=& \frac{1}{4 \pi} \left\{
c_\theta^2 \left[ g \left( \frac{m_h^2}{m_{u1}^2} \right)
+ 2 g \left( \frac{m_{u2}^2}{m_d^2} \right) \right]
+ s_\theta^2 \left[ g \left( \frac{m_h^2}{m_{u2}^2} \right)
  + 2 g \left( \frac{m_{u1}^2}{m_d^2} \right) \right]
\right. \nonumber \\*[1mm] & & \left.
+ 2 g \left( \frac{m_l^2}{m_d^2} \right)
- 9 c_\theta^2 s_\theta^2\, g \left( \frac{m_{u1}^2}{m_{u2}^2} \right) \right\}
%\no & &
+ \frac{1}{2 \pi} \left(
  s_\theta^2 \ln{\frac{m_d^2}{m_{u1}^2}}
  + c _\theta^2 \ln{\frac{m_d^2}{m_{u2}^2}}
  + \ln{\frac{m_d^2}{m_l^2}} \right).
\ea
\es

\section{Generalization}

In this section we generalize the formulas that we derived for leptoquarks.
We consider a set of NP physical scalars with any electric charges,
making the provisos that none of the new scalars has vacuum expectation value
and that they do not mix with the scalars of the SM.
Also,
we assume that all the new scalar fields are complex.

We consider a set of physical New Physics scalars $s$
that have electric charges $Q_s$
(\textit{i.e.}\ each $s$ has a corresponding $Q_s$)
which differ among themselves by integers.\footnote{If
in the NP model under consideration
there are physical scalars with electric charges
that differ among themselves by non-integer values,
then they must be treated separately.
For instance,
a set of NP scalars with electric charges $5/3$,
$2/3$,
$-1/3$,
and $-4/3$
(like the leptoquarks)
must be treated separately from a set of NP scalars
with electric charges $3$,
$2$,
and $1$.}
The charges $Q_s$ may be either positive or negative.

We place all the scalars of each set in a column vector,
ordering the scalars by their decreasing electric charges;
this vector generalizes the vector
\be
\label{fif99}
\left( \begin{array}{c} h \\ u \\ d \\ l \end{array} \right)
\ee
wherein sit the $n_h + n_u + n_d + n_l$ leptoquarks of Section~2.

The column vector of the previous paragraph must constitute a---in general,
reducible---representation of the gauge group $SU(2)$.
Therefore,
the following operators,
represented by matrices,
act on it:
\begin{enumerate}
\item The operator $M = \sqrt{2}\, T_+$,
  where $T_+$ is the raising operator of the gauge group $SU(2)$.
\item The operator $M^\dagger = \sqrt{2}\, T_-$,
  where $T_-$ is the lowering operator of $SU(2)$.
\item The Hermitian operator $\bar M \equiv 2 T_3$,
  where $T_3$ is the third generator of $SU(2)$.
\end{enumerate}
Since the algebra of $SU(2)$ has $T_3 = \left[ T_+, T_- \right]$,
\be
\label{comut}
\bar M = M M^\dagger - M^\dagger M.
\ee
The three operators are,
of course,
represented by matrices that act on the column vector.
The matrix $M$ is off-diagonal:
it only has nonzero matrix elements $M_{ss^\prime}$ for $Q_{s^\prime} = Q_s - 1$.
Conversely,
the nonzero matrix elements $\left( M^\dagger \right)_{ss^\prime}$
only occur for $Q_s = Q_{s^\prime} - 1$.
The matrix $\bar M$ only has nonzero matrix elements $\bar M_{ss^\prime}$
when $Q_{s^\prime} = Q_s$.
Since the scalars in the column vector are ordered
by decreasing electric charges,
one has
\be
M = \left( \begin{array}{ccccccc}
  0 & M_1 & 0 & 0 & \cdots & 0 & 0 \\
  0 & 0 & M_2 & 0 & \cdots & 0 & 0 \\
  0 & 0 & 0 & M_3 & \cdots & 0 & 0 \\
  \vdots & \vdots & \vdots & \vdots & \ddots & \vdots & \vdots \\
  0 & 0 & 0 & 0 & \cdots & 0 & M_n \\
  0 & 0 & 0 & 0 & \cdots & 0 & 0
\end{array} \right).
\label{jvigogfof}
\ee
For instance,
in our leptoquark model we had $M_1 = \N$,
$M_2 = \V$,
and $M_3 = \Q$.
Clearly then,
\be
\bar M = \left( \begin{array}{cccccc}
  M_1 M_1^\dagger & 0 & 0 &  \cdots & 0 & 0 \\
  0 & M_2 M_2^\dagger - M_1^\dagger M_1 & 0 & \cdots & 0 & 0 \\
  \vdots & \vdots & \vdots & \ddots & \vdots & \vdots \\
  0 & 0 & 0 & \cdots & M_n M_n^\dagger - M_{n-1}^\dagger M_{n-1} & 0\\
  0 & 0 & 0 & \cdots & 0 & - M_n^\dagger M_n
\end{array} \right).
\ee
The electric-charge operator is given in this basis by
\be
Q = \left( \begin{array}{cccccc}
  Q_1 \times \mathbbm{1} & 0 & 0 &  \cdots & 0 & 0 \\
  0 & Q_2 \times \mathbbm{1} & 0 & \cdots & 0 & 0 \\
  \vdots & \vdots & \vdots & \ddots & \vdots & \vdots \\
  0 & 0 & 0 & \cdots & Q_n \times \mathbbm{1} & 0\\
  0 & 0 & 0 & \cdots & 0 & Q_{n+1} \times \mathbbm{1}
\end{array} \right),
\label{vnfifo}
\ee
where the unit matrices $\mathbbm{1}$ have the appropriate dimensions,
which may not all be equal.
Note that
\be
Q_1 = Q_2 + 1 = \cdots = Q_n + \left( n - 1 \right) = Q_{n+1} + n.
\ee

One must also take into account the $SU(2)$ commutation relation
$\left[ T_3, T_+ \right] = T_+$.
With $T_3 = \bar M / 2$ and $T_+ = M \left/ \sqrt{2} \right.$,
it yields $\left[ \bar M, M \right] = 2 M$.
Hence,
the submatrices in Eq.~\eqref{jvigogfof} are related by
\bs
\label{5151}
\ba
2 M_1 &=& 2 M_1 M_1^\dagger M_1 - M_1 M_2 M_2^\dagger, \\
2 M_2 &=& 2 M_2 M_2^\dagger M_2 - M_1^\dagger M_1 M_2 - M_2 M_3 M_3^\dagger, \\
\vdots & & \vdots \no
2 M_{n-1} &=& 2 M_{n-1} M_{n-1}^\dagger M_{n-1} - M_{n-2}^\dagger M_{n-2} M_{n-1}
- M_{n-1} M_n M_n^\dagger, \\
2 M_n &=& 2 M_n M_n^\dagger M_n - M_{n-1}^\dagger M_{n-1} M_n.
\ea
\es

The gauge-kinetic Lagrangian for the NP scalars is
\bs
\label{cf93993}
\ba
\mathcal{L} &=& \sum_s
\left( \partial_\mu s + i g s_W A_\mu Q_s s
- i\, \frac{g}{2 c_W}\, Z_\mu \sum_{s^\prime} F_{s s^\prime} s^\prime
\right. \no & & \left.
- i\, \frac{g}{\sqrt{2}}\, W_\mu^+ \sum_{s^\prime} M_{s s^\prime} s^\prime
- i\, \frac{g}{\sqrt{2}}\, W_\mu^- \sum_{s^\prime}
M^\dagger_{s s^\prime} s^\prime \right)
\\ & & \times
\left( \partial^\mu s^\ast - i g s_W A^\mu Q_s s^\ast
+ i\, \frac{g}{2 c_W}\, Z^\mu \sum_{s^{\prime\prime}} F^\ast_{s s^{\prime\prime}}
{s^{\prime\prime}}^\ast
\right. \no & & \left.
+ i\, \frac{g}{\sqrt{2}}\, W^{\mu -} \sum_{s^{\prime \prime}} M^\ast_{s s^{\prime\prime}}
{s^{\prime\prime}}^\ast
+ i\, \frac{g}{\sqrt{2}}\, W^{\mu +} \sum_{s^{\prime\prime}}
M^T_{s s^{\prime\prime}} {s^{\prime\prime}}^\ast \right),
\ea
\es
where
\be
\label{ffff}
F_{s s^\prime} \equiv \bar M_{s s^\prime} - 2 s_W^2 Q_s \delta_{s s^\prime}.
\ee
The Lagrangian~\eqref{cf93993}
produces the following interactions of the scalars with the photons:
\bs
\ba
\mathcal{L}_{ASS} &=& i g s_W A_\theta \sum_s Q_s
\left( s \partial^\theta s^\ast - s^\ast \partial^\theta s \right)
\\
\mathcal{L}_{AASS} &=& g^2 s_W^2 A_\theta A^\theta \sum_s Q_s^2\, s s^\ast.
\ea
\es
It also produces interactions of the scalars with the $W$ gauge bosons,
given by
\bs
\label{www}
\ba
\mathcal{L}_{WSS} &=& i\, \frac{g}{\sqrt{2}}\, \sum_{s, s^\prime} \left[
  \vphantom{\left( M^\dagger \right)_{s s^\prime}}
  W_\theta^+ M_{s s^\prime}
  \left( s^\ast \partial^\theta s^\prime
  - s^\prime \partial^\theta s^\ast \right)
%  \right. \no & & \left.
  + W_\theta^- \left( M^\dagger \right)_{s s^\prime}
  \left( s^\ast \partial^\theta s^\prime
  - s^\prime \partial^\theta s^\ast \right)
\right], \hspace*{7mm}
\\
\mathcal{L}_{WWSS} &=& \frac{g^2}{2}\, W_\theta^+ W^{\theta -}
\sum_{s, s^\prime} \left( M M^\dagger + M^\dagger M \right)_{s s^\prime}
s^\ast s^\prime.
\ea
\es
The interactions of the scalars with the $Z$ are given by
\bs
\label{zzz}
\ba
\mathcal{L}_{ZSS} &=& i\, \frac{g}{2 c_W}\, Z_\theta
\sum_{s, s^\prime} F_{s s^\prime}
\left( s^\ast \partial^\theta s^\prime - s^\prime \partial^\theta s^\ast \right),
\\
\mathcal{L}_{ZZSS} &=& \frac{g^2}{4 c_W^2}\, Z_\theta Z^\theta
\sum_{s, s^\prime} \left( F^2 \right)_{s s^\prime} s^\ast s^\prime.
\ea
\es
Last but not least,
there is the `seagull' interaction of a photon and a $Z$ with two scalars,
given by
\be
\mathcal{L}_{AZSS} = - \frac{g^2 s_W}{c_W}\, A_\theta Z^\theta
\sum_{s, s^\prime} Q_s F_{s s^\prime} s^\ast s^\prime.
\ee

Therefore,
\bs
\ba
A_{\gamma \gamma} \left( q^2 \right) &=&
\frac{g^2 s_W^2}{4 \pi^2}\, \sum_s\, Q_s^2
\left[ B_{00} \left( q^2, m_s^2, m_s^2 \right)
  - \frac{A_0 \left( m_s^2 \right)}{2} \right],
\\
A_{\gamma Z} \left( q^2 \right) &=&
- \frac{g^2 s_W}{8 \pi^2 c_W}\, \sum_s\, Q_s F_{ss}
\left[ B_{00} \left( q^2, m_s^2, m_s^2 \right)
  - \frac{A_0 \left( m_s^2 \right)}{2} \right],
\\
A_{ZZ} \left( q^2 \right) &=&
\frac{g^2}{16 \pi^2 c_W^2} \left[ \sum_{s, s^\prime}
  \left| F_{s s^\prime} \right|^2 B_{00} \left( q^2, m_s^2, m_{s^\prime}^2 \right)
  - \sum_s\left( F^2 \right)_{ss} \frac{A_0 \left( m_s^2 \right)}{2}
  \right],
\\
A_{WW} \left( q^2 \right) &=&
\frac{g^2}{8 \pi^2} \left[ \sum_{s s^\prime}
  \left| M_{s s^\prime} \right|^2 B_{00} \left( q^2, m_s^2, m_{s^\prime}^2 \right)
  - \sum_s \left( M M^\dagger + M^\dagger M \right)_{ss}
  \frac{A_0 \left( m_s^2 \right)}{4}
  \right].
\no & &
\ea
\es
It immediately follows that
\ba
T &=& \frac{1}{2 \pi s_W^2 m_W^2} \left[
  \sum_{s, s^\prime} \left| M_{s s^\prime} \right|^2
  B_{00} \left( 0, m_s^2, m_{s^\prime}^2 \right)
  - \sum_s \left( M M^\dagger + M^\dagger M \right)_{ss}
  \frac{A_0 \left( m_s^2 \right)}{4}
  \right. \no & & \left.
  - \frac{1}{2}\, \sum_{s, s^\prime} \left| F_{s s^\prime} \right|^2
  B_{00} \left( 0, m_s^2, m_{s^\prime}^2 \right)
  + \sum_s \left( F^2 \right)_{ss}
  \frac{A_0 \left( m_s^2 \right)}{4}
  \right],
\ea
where we have used $c_W^2 m_Z^2 = m_W^2$.
Employing Eq.~\eqref{relation},
one obtains
\be
\label{finalt}
T = \frac{1}{16 \pi s_W^2 m_W^2} \left[
  \sum_{s, s^\prime} \left| M_{s s^\prime} \right|^2
  \theta_+ \left( m_s^2, m_{s^\prime}^2 \right)
  - \sum_{s < s^\prime} \left| \bar M_{s s^\prime} \right|^2
  \theta_+ \left( m_s^2, m_{s^\prime}^2 \right)
  \right].
\ee

One furthermore obtains
\bs
\label{finalvw}
\ba
V &=& \frac{1}{8 \pi c_W^2 s_W^2}\, \sum_{s, s^\prime}
\left| F_{s s^\prime} \right|^2 \rho \left( \frac{m_s^2}{m_Z^2},
\frac{m_{s^\prime}^2}{m_Z^2} \right),
\\
W &=& \frac{1}{4 \pi s_W^2}\, \sum_{s, s^\prime}
\left| M_{s s^\prime} \right|^2 \rho \left( \frac{m_s^2}{m_W^2},
\frac{m_{s^\prime}^2}{m_W^2} \right),
\ea
\es
and
\bs
\label{finalx}
\ba
S^{\prime \prime} &=& - \frac{1}{2 \pi}\, \sum_{s, s^\prime}
\left| F_{s s^\prime} \right|^2\, \zeta \left( \frac{m_s^2}{m_Z^2},
\frac{m_{s^\prime}^2}{m_Z^2} \right),
\\
U^{\prime \prime} &=& - S^{\prime \prime} - \frac{1}{\pi}\, \sum_{s, s^\prime}
\left| M_{s s^\prime} \right|^2\, \zeta \left( \frac{m_s^2}{m_W^2},
\frac{m_{s^\prime}^2}{m_W^2} \right),
\\
X &=& - \frac{1}{4 \pi}\, \sum_s Q_s F_{ss}\
\zeta \left( \frac{m_s^2}{m_Z^2}, \frac{m_s^2}{m_Z^2} \right).
\ea
\es

On the other hand,
\be
\label{63}
S^\prime
= \frac{1}{\pi}\, \sum_{s, s^\prime} \left| \bar M_{s s^\prime} \right|^2
B_{00}^\prime \left( 0, m_s^2, m_{s^\prime}^2 \right)
- \frac{2}{\pi}\, \sum_s Q_s \bar M_{ss}\,
B_{00}^\prime \left( 0, m_s^2, m_s^2 \right)
\ee
and
\be
\label{64}
U^\prime = - S^\prime
+ \frac{2}{\pi} \left[ \sum_{s, s^\prime} \left| M_{s s^\prime} \right|^2
B_{00}^\prime \left( 0, m_s^2, m_{s^\prime}^2 \right)
- \sum_s Q_s \bar M_{ss} B_{00}^\prime \left( 0, m_s^2, m_s^2 \right) \right].
\ee

In order to establish the finiteness of $S^\prime$ and of $S^\prime + U^\prime$
we have to make sure that
\bs
\ba
0 &=& \sum_{s, s^\prime} \left| \bar M_{s s^\prime} \right|^2
- 2\, \sum_s Q_s \bar M_{ss},
\label{kvifdo8}
\\
0 &=& \sum_{d, s^\prime} \left| M_{s s^\prime} \right|^2
- \sum_s Q_s \bar M_{ss},
\label{kvifdo9}
\ea
\es
respectively.
Now,
\bs
\ba
\sum_{s, s^\prime} \left| M_{s s^\prime} \right|^2
&=& \mathrm{tr} \left( M M^\dagger \right)
\no &=& \mathrm{tr} \left( M_1 M_1^\dagger + M_2 M_2^\dagger
+ \cdots + M_n M_n^\dagger\right),
\\
\sum_s Q_s \bar M_{ss}
&=& \mathrm{tr} \left( Q \bar M \right)
\no &=& \mathrm{tr} \left[ Q_1 M_1 M_1^\dagger
  + Q_2 \left( M_2 M_2^\dagger - M_1^\dagger M_1 \right)
  + \cdots
  \right. \no & & \left.
  + Q_n \left( M_n M_n^\dagger - M_{n-1}^\dagger M_{n-1} \right)
  - Q_{n+1} M_n^\dagger M_n \right]
\no &=& \mathrm{tr} \left[ \left( Q_1 - Q_2 \right) M_1 M_1^\dagger
  + \left( Q_ 2 - Q_3 \right) M_2 M_2^\dagger
  + \cdots
  \right. \no & & \left.
  + \left( Q_n - Q_{n+1} \right) M_n M_n^\dagger\right],
\label{66b}
\ea
\es
which proves Eq.~\eqref{kvifdo9},
because $Q_1 - Q_2 = Q_2 - Q_3 = \cdots = Q_n - Q_{n+1} = 1$.
Moreover,
\bs
\ba
\sum_{s, s^\prime} \left| \bar M_{s s^\prime} \right|^2
&=& \mathrm{tr} \left( \bar M^2 \right)
\\ &=& \mathrm{tr} \left[ \left( M_1 M_1^\dagger \right)^2
  + \left( M_2 M_2^\dagger - M_1^\dagger M_1 \right)^2
  + \left( M_3 M_3^\dagger - M_2^\dagger M_2 \right)^2
  + \cdots
  \right. \no & & \left.
  + \left( M_n M_n^\dagger - M_{n-1}^\dagger M_{n-1} \right)^2
  + \left( - M_n^\dagger M_n \right)^2 \right]
\\ &=& 2\, \mathrm{tr} \left( M_1 M_1^\dagger + M_2 M_2^\dagger + M_3 M_3^\dagger
+ \cdots + M_{n-1} M_{n-1}^\dagger + M_n M_n^\dagger \right),
\ea
\es
because of Eqs.~\eqref{5151}.
Therefore,
using Eq.~\eqref{66b},
we see that Eq.~\eqref{kvifdo8} is also true.

Employing Eq.~\eqref{divv} on Eqs.~\eqref{63} and~\eqref{64},
one at last obtains
\bs
\label{finalsu}
\ba
S^\prime &=& \frac{1}{24 \pi} \left\{ \sum_{s, s^\prime}
\left| \bar M_{s s^\prime} \right|^2
g \left( \frac{m_s^2}{m_{s^\prime}^2} \right)
+ 2\, \sum_s
\left[ \left( \bar M^2 \right)_{ss}
- 2 Q_s \bar M_{ss} \right] \ln{m_s^2} \right\},
\\
U^\prime &=&
- S^\prime + \frac{1}{12 \pi} \left\{ \sum_{s, s^\prime}
\left| M_{s s^\prime} \right|^2 g \left( \frac{m_s^2}{m_{s^\prime}^2} \right)
+ \sum_s \left[
  \left( M M^\dagger + M^\dagger M \right)_{ss} - 2 Q_s \bar M_{ss} \right]
\ln{m_s^2} \right\}.
\no & &
\ea
\es

\section{Conclusions}

In this paper we have derived general formulas for the oblique parameters
in a model with an arbitrary number of scalar leptoquarks.
Those formulas include the mixing matrices $\N$,
$\V$,
and $\Q$ that appear in the charged current of Eq.~\eqref{Wphiphi};
and also the matrices $\bar H$,
$\bar U$,
$\bar D$,
and $\bar L$ that are derived from $\N$,
$\V$,
and $\Q$ through Eqs.~\eqref{bars} and~\eqref{vcvufgio}.
The final formulas for the oblique parameters are $T$ given by Eq.~\eqref{t2},
$V$ and $W$ given by Eqs.~\eqref{33},
$X$ given by Eq.~\eqref{XXXXX},
and $S$ and $U$ given by Eqs.~\eqref{tttt1} and~\eqref{t5}.
The relevant functions $g$,
$\theta_+$,
$\rho$,
and $\zeta$ are given by Eqs.~\eqref{functiong},
\eqref{theta+},
\eqref{681},
and~\eqref{zeta1},
respectively.

We have then considered the more general situation
where the New Physics is constituted by (complex) physical scalars $s$
that originate in arbitrary representations of the gauge group;
those scalars are allowed to freely mix among themselves,
but not to mix with the scalar doublet of the SM.
In this case the fundamental mixing matrix is $M$
that appears in Eq.~\eqref{www};
therefrom one derives $\bar M$ in Eq.~\eqref{comut}
and $F$ in Eq.~\eqref{ffff}.
The final results for the oblique parameters are then given
in Eqs.~\eqref{finalt},
\eqref{finalvw},
\eqref{finalx},
and~\eqref{finalsu}.

\vspace*{5mm}

\paragraph{Acknowledgements:}
The authors thank the Portuguese Foundation for Science and Technology
for support through the projects UIDB/00777/2020,
UIDP/00777/2020,
and CERN/FIS-PAR/0002/2021.
The work of F.A.\ was furthermore supported by grant UI/BD/153763/2022.
The work of L.L.\ was furthermore supported by project CERN/FIS-PAR/0019/2021.

\newpage

\begin{appendix}

\section{Parameter counting}

\setcounter{equation}{0}
\renewcommand{\theequation}{A\arabic{equation}}

In our leptoquark models the total number of parameters is
$n_\mathrm{total} = n_\mathrm{masses} + n_\mathrm{mixings}$.
Here,
\ba
n_\mathrm{masses} &=& n_h + n_u + n_d + n_l
\no &=& n_{\delta,7}
+ \left( n_{\delta,7} + n_{\delta,1} + n_{\tau,-2} \right)
+ \left( n_{\sigma,-2} + n_{\delta,1} + n_{\tau,-2} \right)
+ \left( n_{\sigma,-8} + n_{\tau,-2} \right) \hspace*{5mm}
\label{nmasses}
\ea
is the total number of distinct leptoquark masses;
$n_\mathrm{mixings}$ is the number of independent parameters
in the mixing matrices.
The latter may be computed as follows.

Since the matrices $\Hh$,
$\Uu$,
$\Dd$,
and $\Ll$ may be derived fom the matrices $\N$,
$\V$,
and $\Q$ through Eqs.~\eqref{bars} and~\eqref{vcvufgio},
in order to know $n_\mathrm{mixings}$ one must count the parameters
in the latter matrices.
They are derived from $H_1$,
$U_1$,
$U_2$,
$U_3$,
$D_2$,
$D_3$,
and $L_2$ through Eqs.~\eqref{NVQ}.
Let us assume,
for the sake of simplicity,
the latter matrices to be real.
We use the fact that the first $m$ rows of an $n \times n$ real orthogonal
matrix may be parameterized
by $\left. m \left( 2 n - m - 1 \right) \right/ 2$ parameters.
Therefore,
\bs
\ba
H_1
& \mbox{has} &
\frac{n_{\delta,7} \left( n_{\delta,7} - 1 \right)}{2}\, \
\mbox{parameters},
\\
\left( \begin{array}{c} U_1 \\ U_2 \\ U_3 \end{array} \right)
& \mbox{has} &
\frac{\left( n_{\delta,7} + n_{\delta,1} + n_{\tau,-2} \right)
  \left( n_{\delta,7} + n_{\delta,1} + n_{\tau,-2} - 1 \right)}{2}\, \
\mbox{parameters},
\label{dvifofd}
\\
\left( \begin{array}{c} D_2 \\ D_3 \end{array} \right)
& \mbox{has} &
\frac{\left( n_{\delta,1} + n_{\tau,-2} \right)
  \left( 2 n_{\sigma,-2} + n_{\delta,1} + n_{\tau,-2} - 1 \right)}{2}\, \
\mbox{parameters},
\\
L_2
& \mbox{has} &
\frac{n_{\tau,-2} \left( 2 n_{\sigma,-8} + n_{\tau,-2} - 1 \right)}{2}\, \
\mbox{parameters}.
\ea
\es

However,
one must take into account the fact that,
in Eqs.~\eqref{NVQ},
the transformations
\bs
\ba
& & H_1 \to A H_1, \ U_1 \to A U_1,
\\
& & U_2 \to B U_2, \ D_2 \to B D_2,
\\
& & U_3 \to C U_3, \ D_3 \to C D_3, \ L_2 \to C L_2,
\ea
\es
where $A$,
$B$,
and $C$ are arbitrary $n_{\delta,7} \times n_{\delta,7}$,
$n_{\delta,1} \times n_{\delta,1}$,
and $n_{\tau,-2} \times n_{\tau,-2}$ real orthogonal matrices,
respectively,
leave $\N$,
$\V$,
and $\Q$ invariant.
Therefore
the total number of parameters in $\N$,
$\V$,
and $\Q$ is
\ba
n_\mathrm{mixings} &=& \frac{n_{\delta,7} \left( n_{\delta,7} - 1 \right)}{2}
+ \frac{\left( n_{\delta,7} + n_{\delta,1} + n_{\tau,-2} \right)
  \left( n_{\delta,7} + n_{\delta,1} + n_{\tau,-2} - 1 \right)}{2}
\no & &
+ \frac{\left( n_{\delta,1} + n_{\tau,-2} \right)
  \left( 2 n_{\sigma,-2} + n_{\delta,1} + n_{\tau,-2} - 1 \right)}{2}
+ \frac{n_{\tau,-2} \left( 2 n_{\sigma,-8} + n_{\tau,-2} - 1 \right)}{2}
\no & & - \frac{n_{\delta,7} \left( n_{\delta,7} - 1 \right)}{2}
- \frac{n_{\delta,1} \left( n_{\delta,1} - 1 \right)}{2}
- \frac{n_{\tau,-2} \left( n_{\tau,-2} - 1 \right)}{2}.
\label{nmixings}
\ea

Utilizing both Eqs.~\eqref{nmasses} and~\eqref{nmixings},
we find that the leptoquark models with the smallest numbers of parameters
are the ones in Table~\ref{table1}.
\begin{table}[ht]
  \centering
    \begin{tabular}{|ccccc|cccc|ccc|}
    \hline
    $n_{\sigma,-2}$ & $n_{\sigma,-8}$ & $n_{\delta,1}$ & $n_{\delta, 7}$ &
    $n_{\tau,-2}$ & $n_h$  & $n_u$ & $n_d$ & $n_l$ &
    $n_\mathrm{masses}$ & $n_\mathrm{mixings}$ & $n_\mathrm{total}$ \\
    \hline
    0     & 0     & 1     & 0     & 0     & 1     & 1     & 0     & 0     & 2     & 0     & 2 \\
    0     & 0     & 0     & 1     & 0     & 0     & 1     & 1     & 0     & 2     & 0     & 2 \\ \hline
    0     & 0     & 0     & 0     & 1     & 0     & 1     & 1     & 1     & 3     & 0     & 3 \\
    1     & 0     & 1     & 0     & 0     & 1     & 1     & 1     & 0     & 3     & 0     & 3 \\
    0     & 1     & 1     & 0     & 0     & 1     & 1     & 0     & 1     & 3     & 0     & 3 \\
    0     & 1     & 0     & 1     & 0     & 0     & 1     & 1     & 1     & 3     & 0     & 3 \\ \hline
    1     & 0     & 0     & 1     & 0     & 0     & 1     & 2     & 0     & 3     & 1     & 4 \\
    2     & 0     & 1     & 0     & 0     & 1     & 1     & 2     & 0     & 4     & 0     & 4 \\
    0     & 2     & 1     & 0     & 0     & 1     & 1     & 0     & 2     & 4     & 0     & 4 \\
    1     & 1     & 1     & 0     & 0     & 1     & 1     & 1     & 1     & 4     & 0     & 4 \\
    0     & 2     & 0     & 1     & 0     & 0     & 1     & 1     & 2     & 4     & 0     & 4 \\ \hline
    0     & 1     & 0     & 0     & 1     & 0     & 1     & 1     & 2     & 4     & 1     & 5 \\
    1     & 0     & 0     & 0     & 1     & 0     & 1     & 2     & 1     & 4     & 1     & 5 \\
    0     & 0     & 1     & 1     & 0     & 1     & 2     & 1     & 0     & 4     & 1     & 5 \\
    1     & 1     & 0     & 1     & 0     & 0     & 1     & 2     & 1     & 4     & 1     & 5 \\
    0     & 0     & 2     & 0     & 0     & 2     & 2     & 0     & 0     & 4     & 1     & 5 \\
    0     & 0     & 0     & 2     & 0     & 0     & 2     & 2     & 0     & 4     & 1     & 5 \\
    3     & 0     & 1     & 0     & 0     & 1     & 1     & 3     & 0     & 5     & 0     & 5 \\
    0     & 3     & 1     & 0     & 0     & 1     & 1     & 0     & 3     & 5     & 0     & 5 \\
    2     & 1     & 1     & 0     & 0     & 1     & 1     & 2     & 1     & 5     & 0     & 5 \\
    1     & 2     & 1     & 0     & 0     & 1     & 1     & 1     & 2     & 5     & 0     & 5 \\
    0     & 3     & 0     & 1     & 0     & 0     & 1     & 1     & 3     & 5     & 0     & 5 \\
    \hline
    \end{tabular}
    \caption{The leptoquark models with the smallest numbers of parameters.
      Models with $n_\mathrm{total} > 5$
      and (trivial) models with $n_{\delta,1} = n_{\delta, 7} = n_{\tau, -2} = 0$
      are not included.
    \label{table1}}
\end{table}

\section{Parameterization of the mixing}

\setcounter{equation}{0}
\renewcommand{\theequation}{B\arabic{equation}}

One may choose to parameterize the mixing matrices in the following way:
\begin{enumerate}
\item One sets $H_1 = \mathbbm{1}_{n_{\delta,7}}$.
\item One leaves the matrix
  \be
  \left( \begin{array}{c} U_1 \\ U_2 \\ U_3 \end{array} \right)
  \ee
  fully free,
  \textit{i.e.}\ with all its parameters.
\item One parameterizes the matrix
  \be
  \left( \begin{array}{c} L_1 \\ L_2 \end{array} \right)
  \label{ioe00}
  \ee
  through $\left. n_{\tau,-2} \left( n_{\tau,-2} - 1 \right) \right/ 2$
  rotations among the rows of $L_2$
  followed by $n_{\sigma,-8} n_{\tau,-2}$ rotations
  between the rows of $L_1$ and the ones of $L_2$.
  For instance,
  if $n_{\sigma,-8} = n_{\tau,-2} = 2$,
  one parameterizes the matrix~\eqref{ioe00} as
  $O_{34} O_{13} O_{14} O_{23} O_{24}$.
\item One parameterizes the matrix
  \be
  \left( \begin{array}{c} D_1 \\ D_2 \\ D_3 \end{array} \right)
  \label{ioe11}
  \ee
  through $n_{\sigma,-2} n_{\delta,1}$ rotations
  between the rows of $D_1$ and the ones of $D_2$,
  followed by $n_{\sigma,-2} n_{\tau,-2}$ rotations
  between the rows of $D_1$ and the ones of $D_3$
  and by $n_{\delta,1} n_{\tau,-2}$ rotations
  between the rows of $D_2$ and the ones of $D_3$.
  For instance,
  if $n_{\sigma,-2} = n_{\delta,1} = 2$ and $n_{\tau,-2} = 1$,
  one parameterizes the matrix~\eqref{ioe11} as
  $O_{13} O_{14} O_{23} O_{24} O_{15} O_{25} O_{35} O_{45}$.
\end{enumerate}

\section{Cancellation of the divergences in $S^\prime$ and $U^\prime$}

\setcounter{equation}{0}
\renewcommand{\theequation}{C\arabic{equation}}

Because of Eqs.~\eqref{jgfogh},
\bs
\label{bjbgig}
\ba
n_h &=& n_{\delta,7}, \\
\mathrm{tr} \left( U_1^\dagger U_1 \right) &=& n_{\delta,7}, \\
\mathrm{tr} \left( U_2^\dagger U_2 \right) &=& n_{\delta,1}, \\
\mathrm{tr} \left( U_3^\dagger U_3 \right) &=& n_{\tau,-2}, \\
\mathrm{tr} \left( D_2^\dagger D_2 \right) &=& n_{\delta,1}, \\
\mathrm{tr} \left( D_3^\dagger D_3 \right) &=& n_{\tau,-2}, \\
\mathrm{tr} \left( L_2^\dagger L_2 \right) &=& n_{\tau,-2}.
\ea
\es
Using Eqs.~\eqref{NVQ} and the unitarity of the matrices in Eq.~\eqref{unit},
the matrices~\eqref{vcvufgio} may be rewritten
\bs
\ba
\Hh &=& \mathbbm{1}_{n_h},
\\
\Uu &=& - U_1^\dagger U_1 + U_2^\dagger U_2 + 2\, U_3^\dagger U_3,
\\
\Dd &=& D_2^\dagger D_2,
\\
\Ll &=& 2\, L_2^\dagger L_2.
\ea
\es

\paragraph{Divergences:} When $\epsilon \to 0^+$,
the quantity
\be
\label{div}
\mathrm{div} \equiv \frac{2}{\epsilon}
- \gamma + \ln{\left( 4 \pi \right)}
\ee
diverges.
In Eq.~\eqref{div},
$\gamma$ is Euler's constant.
The PV function
\be
B_{00} \left( q^2, m_1^2, m_2^2 \right) =
\left( \frac{m_1^2 + m_2^2}{4} - \frac{q^2}{12} \right) \mathrm{div}
+ \mathrm{convergent\ terms}.
\label{div1}
\ee
Therefore,
\bs
\label{divB}
\ba
B_{00}^\prime \left( q^2, m_1^2, m_2^2 \right) &=& - \frac{\mathrm{div}}{12}
+ \mathrm{convergent\ terms},
\\
\bar B_{00} \left( q^2, m_1^2, m_2^2 \right) &=& - \frac{\mathrm{div}}{12}
+ \mathrm{convergent\ terms}.
\ea
\es

\paragraph{$S^\prime$ divergence:} Taking into account Eqs.~\eqref{divB},
the divergent part of $S^\prime$ in Eq.~\eqref{s3} is proportional to
\ba
& &
\mathrm{tr}\, \Hh^2 - 2 Q_h\, \mathrm{tr}\, \Hh
+ \mathrm{tr}\, \Uu^2 - 2 Q_u\, \mathrm{tr}\, \Uu
+ \mathrm{tr}\, \Dd^2 + 2 Q_d\, \mathrm{tr}\, \Dd
+ \mathrm{tr}\, \Ll^2 + 2 Q_l\, \mathrm{tr}\, \Ll
\no &=&
\mathrm{tr} \left( \mathbbm{1}_{n_h} \right)
- 2 Q_h\, \mathrm{tr} \left( \mathbbm{1}_{n_h} \right)
\no & &
+ \mathrm{tr} \left( U_1^\dagger U_1 + U_2^\dagger U_2
+ 4 U_3^\dagger U_3 \right)
- 2 Q_u\, \mathrm{tr} \left( - U_1^\dagger U_1 + U_2^\dagger U_2
+ 2 U_3^\dagger U_3 \right)
\no & &
+ \mathrm{tr} \left( D_2^\dagger D_2 \right)
+ 2 Q_d\, \mathrm{tr} \left( D_2^\dagger D_2 \right)
\no & &
+ 4\, \mathrm{tr} \left( L_2^\dagger L_2 \right)
+ 4 Q_l\, \mathrm{tr} \left( L_2^\dagger L_2 \right)
\no &=&
n_{\delta,7} - 2 Q_h n_{\delta,7}
\no & &
+ n_{\delta,7} + n_{\delta,1} + 4 n_{\tau,-2}
- 2 Q_u \left( - n_{\delta,7} + n_{\delta,1} + 2 n_{\tau,-2} \right)
\no & &
+ n_{\delta,1} + 2 Q_d n_{\delta,1}
\no & &
+ 4 n_{\tau,-2} + 4 Q_l n_{\tau,-2}
\no &=&
n_{\delta,7} \left( 2 - 2 Q_h + 2 Q_u \right)
+ n_{\delta,1} \left( 2 - 2 Q_u + 2 Q_d \right)
+ n_{\tau,-2} \left( 8 - 4 Q_u + 4 Q_l \right)
\no &=& 0,
\ea
where we have used Eqs.~\eqref{bjbgig}.
Thus,
the parameter $S^\prime$ is finite.

\paragraph{$S^\prime + U^\prime$ divergence:} The divergent part
of $S^\prime + U^\prime$ in Eq.~\eqref{u4}
is proportional to
\ba
& &
\mathrm{tr} \left( \N \N^\dagger + \V \V^\dagger + \Q \Q^\dagger \right) 
- \left( Q_h\, \mathrm{tr}\, \Hh + Q_u\, \mathrm{tr}\, \Uu \right)
+ \left( Q_d\, \mathrm{tr}\, \Dd + Q_l\, \mathrm{tr}\, \Ll \right)
\no &=&
\mathrm{tr} \left( \N \N^\dagger + \V \V^\dagger + \Q \Q^\dagger \right) 
- Q_h\, \mathrm{tr} \left( \N \N^\dagger \right)
- Q_u\, \mathrm{tr} \left( \V \V^\dagger - \N \N^\dagger \right)
\no & &
+ Q_d\, \mathrm{tr} \left( \V \V^\dagger - \Q \Q^\dagger \right)
+ Q_l\, \mathrm{tr} \left( \Q \Q^\dagger \right)
\no &=&
\left( 1 - Q_h + Q_u \right) \mathrm{tr} \left( \N \N^\dagger \right)
+ \left( 1 - Q_u + Q_d \right) \mathrm{tr} \left( \V \V^\dagger \right)
+ \left( 1 - Q_d + Q_l \right) \mathrm{tr} \left( \Q \Q^\dagger \right)
\no &=& 0,
\ea
where we have used $Q_h - Q_u = Q_u - Q_d = Q_d - Q_l = 1$.
Thus,
$S^\prime + U^\prime$ is finite.

\end{appendix}

\end{fmffile}

\end{document}